\documentclass[american,twocolumn, nofootinbib, nobalancelastpage,superscriptaddress]{revtex4-2}
\usepackage[T1]{fontenc}
\usepackage[utf8]{inputenc}
\usepackage{babel}
\usepackage{amsmath}
\usepackage{graphicx}
\usepackage[]
 {hyperref}

\usepackage{amssymb,bm,amsthm,mathtools}

\makeatletter

\usepackage{quiver}  %
\usetikzlibrary{decorations.pathmorphing, calc}

\makeatletter
\renewcommand*{\fnum@figure}{{\normalfont\bfseries \figurename~\thefigure}}
\renewcommand*{\@caption@fignum@sep}{\textbf{: }}
\makeatother

\usepackage[noindentafter]{titlesec}

\titleformat{\section}{\raggedright\bfseries\large}{\Roman{section}.}{0.3em}{}
\titlespacing\section{0pt}{8pt plus 3pt minus 2pt}{3pt} %

\titleformat{\subsection}{\raggedright\bfseries}{\Roman{section}.\Alph{subsection}}{1em}{}

\renewcommand*\citet[1]{\cite{#1}}

\makeatletter
\g@addto@macro\bfseries{\boldmath}
\makeatother

\usepackage{amssymb,bm,amsthm}

\RequirePackage{xcolor}
\definecolor{darkgreen}{rgb}{0, 0.3, 0}
\definecolor{midgreen}{rgb}{0.5, 0.8, 0.5}
\definecolor{darkred}{rgb}{0.6, 0, 0}
\definecolor{darkblue}{rgb}{0, 0, 0.6}
\definecolor{darkorange}{rgb}{0.5, 0.2, 0}

\newcommand\red[1]{\textcolor{darkred}{#1}}
\newcommand\blue[1]{\textcolor{darkblue}{#1}}
\newcommand\green[1]{\textcolor{darkgreen}{#1}}
\newcommand\orange[1]{\textcolor{darkorange}{#1}}

\RequirePackage{hyperref}
\hypersetup{linktocpage=true, colorlinks=true, citecolor={darkgreen}, linkcolor={darkred}, urlcolor={darkblue} }

\newcommand{\refsub}[2]{\hyperref[#1]{\ref*{#1}#2}}

\newcommand{\physics}{Department of Physics and Quantitative Biology Institute, Yale University, New Haven, CT, USA}

\newcommand{\mcdb}{Molecular, Cellular, and
Developmental Biology, Yale University, New Haven, CT, USA}

\makeatother

\begin{document}
\title{Discrete turn strategies emerge in information-limited navigation}
\date{June 22, 2026}
\author{Jose~M.~Betancourt}
\affiliation{\physics}
\author{Matthew~P.~Leighton}
\affiliation{\physics}
\author{Thierry~Emonet}
\affiliation{\physics}
\affiliation{\mcdb}
\author{Benjamin~B.~Machta}
\email{benjamin.machta@yale.edu} 
\affiliation{\physics}
\author{Michael~C.~Abbott}
\email{michael.abbott@yale.edu} 
\affiliation{\physics}

\begin{abstract}
\noindent 
Navigation up a smooth sensory gradient is one of the simplest behavioural tasks, and some organisms solve it by making continuous adjustments to their course. 
Bacteria instead employ a variety of discrete strategies, including run and tumble motion, direction reversals, and turns by specific angles.
Here we ask what drives the choice of these strategies, framing the problem as maximising up-gradient speed with a given amount of sensory information per unit time. 
We find that, without directional information on which way to turn, behavioural strategies that take discrete actions perform better than gradual steering. 
As the amount of information is increased, we see a series of transitions between optimal strategies, including a shift from direction reversals to fully re-orienting tumbles.
Among more complex re-orientation strategies, we show that discrete turn angles are best, and observe transitions in the number of angles employed by the optimal strategy. 
More broadly, such emergent simplicity in behaviour is a tractable example of a widespread phenomenon in which biology chooses a discrete solution, despite the underlying physics being continuous.
\end{abstract}
\maketitle

\section*{Introduction}

Many micro-organisms navigate by switching between a small set of
discrete behavioural states in response to continuous changes in their
sensory input. Run and tumble is one such strategy, alternating between
swimming as straight as possible, and suddenly picking a new direction
at random \citet{berg1975how}. Other organisms swim backwards and
forwards \citet{taylor1974reversal}, or make right-angle turns \citet{xie2011bacterial}.
Why do they do these things, instead of making continuous changes?
And how do they choose which strategy to adopt? In this paper we offer an explanation based only on the constraint that they have limited sensory information.

These questions about navigation belong to a wide class of related questions about why biology chooses discrete solutions where the underlying physics is continuous, or nearly so. For example, the human tongue can make a continuum of vowel sounds, but every language discretises this space to give meaning to a small number of phonemes. Why do this, and what controls the number? 
In the salamander, thousands of retinal ganglion cells have the thresholds at which they turn on or off set to just three values \cite{kastner2015critical}. Why not use intermediate settings?
More microscopically, most ion channels are either open or closed. Why not allow a graded current through the membrane? None of these questions are at the atomic scale, where the ultimate discreteness of the underlying physics would enter. 

What all of these systems have in common is that they involve information processing.
Mathematically, optimising an objective function containing a measure of information often leads to a discrete solution \cite{smith1971information,sims2006rational,jung2019discrete,mattingly2018maximizing,abbott2019scaling,wang2026improved}.
If evolution has performed optimisation, then perhaps the simple discrete solutions seen in its products are in a sense deliberate. %
Perhaps they aren't compromises based on the difficulty of making control systems out of proteins, nor the result of a trade-off between performance and complexity, but rather, signs that something close to the ideal solution has been attained. Getting access to these these ideas about simplicity, and what we term emergent discreteness, is our main motivation for studying toy models of a behavioural task. 

The task we study is navigation up a fixed sensory gradient.
This task is solved by many organisms, and the case of \emph{E. coli} in a shallow gradient of attractant 
has been shown to be close to optimal at using the information it does receive to execute its chosen strategy~\cite{mattingly2021escherichia}. In this bacterium, sensory proteins bind to attractant molecules, and these binding events lead to changes in its internal chemical state, which in turn regulates the motors that drive the flagella~\cite{karmakar2021state}. The noise in every step of this process limits how much information flows from heading to behaviour. We focus on this end-to-end information rate, to obtain a framework in which to study how limited information shapes behavioural strategies, while remaining agnostic to organism-specific details.

However, while bacteria can typically only measure how the concentration changes along their path, larger organisms can directly measure spatial gradients across their bodies. For example, fruit flies compare signals between their two antennae, allowing them to measure the instantaneous gradient perpendicular to their line of travel. It will be important to make a distinction between such directional information and the non-directional kind collected by \emph{E. coli}. While both may be expressed in bits per second, knowing which way to steer is qualitatively different, and changes what strategies are preferred.

What we find is that if the sensor does not provide directional information,
then continuous steering is always worse than sudden changes of direction. As the
amount of sensory information varies, for instance because the gradient
steepness changes, we see various transitions of which strategy is
best. Further, if we allow arbitrary actions, then the optimal strategy
employs a discrete set of turn angles. These results are an example
of how discreteness emerges from optimisation with limited information,
a phenomenon which may explain many otherwise surprising discrete
actions in biology.

    \begin{figure}
    \includegraphics[width=1\columnwidth]{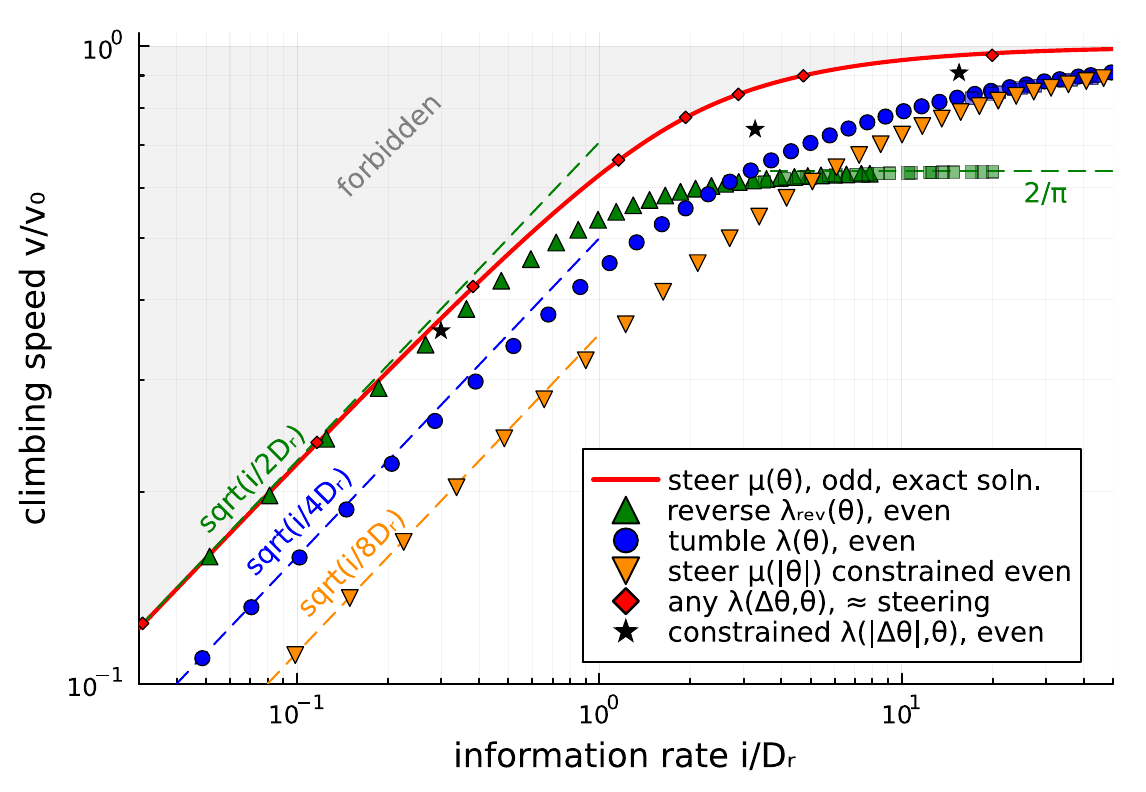}
    
    \caption{Performance of strategies for two-dimensional navigation, on the speed-information
    plane. The continuous steering strategy achieves the highest performance,
    provided the sign of heading $\theta$ is visible (red). Among strategies
    not using such directional information, i.e. those described by even
    functions of $\theta$, fully re-orienting instantaneous tumbles (blue)
    perform well at high information rates, but at lower rates reversing
    (green) needs half as much information for the same speed. All low-information
    limits scale as $v\sim\sqrt{i}$, and high-information limits are
    $v/v_{0}\to1$ except reverse, which is limited to $v/v_{0}<2/\pi\approx0.64$.
    Information is measured in nats, thus the axis is in nats per rotational
    diffusion time. Plot points are numerical, and square points use ansatz $\lambda_{\mathrm{strong}}(\theta)$. Black stars
    are the three solutions shown in figure \ref{fig:Any-jump}.}
    \label{fig:Performance-2D}
    \end{figure}

\section*{Results}

For simplicity we start with navigation in two dimensions, with an
agent swimming at some fixed speed $v_{0}$, at an angle $\theta$ away from the desired direction. Its mean speed towards the goal is then $v=v_{0}\left\langle \cos\theta\right\rangle _{\theta}$.
The angle $\theta$ suffers rotational diffusion $D_{r}$, which tends to randomise
the heading in time of order $1/D_{r}$. To make progress, this passive noise from diffusion
must be counteracted by a control system. 

We assume the the agent senses some information about its instantaneous heading, and uses this to produce a controlled action $d\theta_c$. The active and passive contributions are added to obtain the full update $d\theta$ for time $dt$. Our system can be sketched as follows:
\[\begin{tikzpicture}[>=stealth, scale=0.75]
  \node (state)  at (0,1) {$\text{state } \theta$};
  \node (diff)   at (5.7,1) {$\text{diffusion } D_r$};
  \node (signal) at (3,0) {$\text{signal}$};
  \node (action) at (6,0) {$\text{action } d\theta_c$};
  \node (update) at (9,0) {$\text{update } d\theta$};

  \draw[->] (state) -- node[above right] {$s$} (signal); 
  \draw[->] (diff)  -- (update);                         
  \draw[->] (signal) -- node[above] {$c$} (action);
  \draw[->] (action) -- (update);

  \draw[dashed] (state.south)
      -- (state.south |- {$(action.south)+(0,-0.4)$})
      -- node[below] {$\text{information rate } i$}
         (action.south |- {$(action.south)+(0,-0.4)$})
      -- (action.south);
\end{tikzpicture}\]
We measure the fidelity of the control response by the mutual information rate from state to action, $i=I(\Theta;d\Theta_{c})/dt$.
In any real system both sensing and control are noisy, and both place limits on the response. We do not model them separately, and describe a strategy by some function of heading $\theta$ with Poisson or diffusive noise.
However, to study strategies available without directional sensing, we will impose that the response be the same for $\theta$ and $-\theta$. This can be viewed as inserting $s(\theta)=\lvert\theta\rvert$ while optimising over $c$. (We also treat $s(\theta)=\mathop{\mathrm{sign}}(\theta)$ in appendix \ref{sec:signonly}.)

To find the control strategy
which maximises mean climbing speed at a given information rate,
we introduce a trade-off parameter $\gamma>0$, to arrive at the following optimisation problem: 
\begin{equation}
\max_{c}\big(v/v_{0}-\gamma\:i\big).\label{eq:objective}
\end{equation}
Finding the optimal control strategy $c$ for many different values
of $\gamma$ gives us a frontier on the speed-information plane (figure \ref{fig:Performance-2D}). 
We do this first for some particular classes of controller,
before turning to the general case of an arbitrary instantaneous response.

    \begin{figure*}
    \includegraphics[width=1\textwidth]{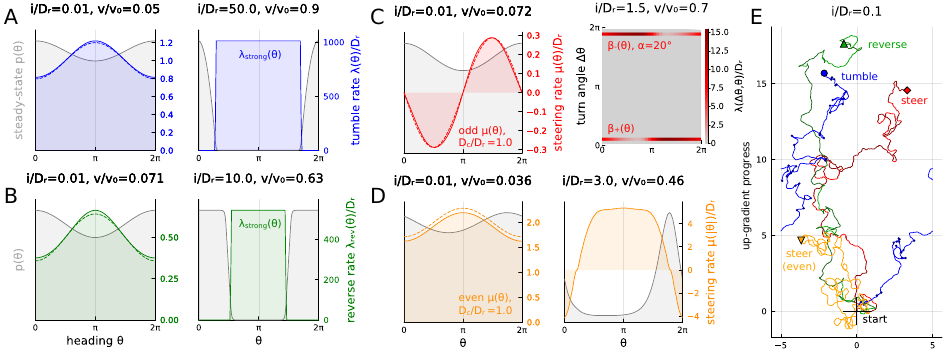}
    
    \caption{Numerical optimal strategies for run \& tumble, steering, and reversing.
    (A) Tumble rate $\lambda(\theta)$ in blue, and steady-state $p(\theta)$
    in grey, for a low-information case, and a high-information case using
    $\lambda_{\mathrm{strong}}(\theta)$. (B) The run-reverse strategy achieves
    $\sqrt{2}$ times the speed at similar information rate $i/D_{r}\approx0.01$,
    by acting at half the rate of the tumble strategy. But its high-information
    case saturates at speed $2/\pi$, when $p(\theta)$ is uniform on
    $-\pi/2\protect\leq\theta\protect\leq\pi/2$. (C) Steering rate $\mu(\theta)$
    in red, and corresponding $p(\theta)$ in grey. Second panel translates turn rate $\mu(\theta)$
    and controller noise $D_{c}=D_{r}$ to Poisson rate $\lambda(\Delta\theta,\theta)$
    using two small angles $\Delta\theta=\pm\alpha$. (D) Steering rate $\mu(\left|\theta\right|)$,
    assuming the sign of $\theta$ is not observable. At low information
    rate, $i/D_{r}\approx0.01$, the agent always turns in one direction
    but modulates its turning speed. At high information rate, this strategy
    creates stable \& unstable fixed points at some $\pm\theta_{\star}$.
    (A-D) Dashed lines are analytic results at low information, from the
    text. (E) Sample trajectories for four strategies, all with $i/D_{r}\approx0.1$
    (times $0<t<50$ in units $D_{r}=v_{0}=1$, wrapped to $-5<x<5$).
    More trajectories are shown in  appendix \ref{sec:more-figures}.}
    \label{fig:Numerical-2D}
    \end{figure*}

\section*{Steering vs. tumbling, in two dimensions}

\label{sec:Steering-vs-tumbling}

We begin by comparing two behaviours: \emph{steering}, which continuously
updates the heading at rate $\mu$, and \emph{tumbling}, which picks
a new heading at random with rate $\lambda$. We divide the evolution
of the heading $d\theta$ in time $dt$ into a controlled part $d\theta_{c}$
and Gaussian noise $dW$ from diffusion: 
\begin{equation}
\begin{aligned}d\theta & =d\theta_{c}+\sqrt{2D_{r}}dW\\
 & \quad\;d\theta_{c}=\red{\mu(\theta)\,dt}+\blue{\Delta\theta\,dJ(\lambda(\theta))}+\red{\sqrt{2D_{c}}dW'}.
\end{aligned}
\label{eq:dtheta}
\end{equation}
Here $dJ$ is a Poisson jump process, $\Delta\theta\sim U(-\pi,\pi)$
is the change in heading after a tumble, and $D_{c}$ is controller
noise. The equivalent Fokker-Planck equation is 
\begin{align*}
 & \frac{dp(\theta)}{dt}=\blue{-\lambda(\theta)p(\theta)+\int\negmedspace\frac{d\theta}{2\pi}p(\theta)\lambda(\theta)}\quad\text{(tumbling)}\\
 & \quad\red{-\frac{d\left[\mu(\theta)p(\theta)\right]}{d\theta}}+(\red{D_{c}}+D_{r})\frac{d^{2}p(\theta)}{d\theta^{2}}\quad\text{(steering \& noise)}.
\end{align*}
Given the description of the strategy by $\lambda(\theta)$, $\mu(\theta)$
and $D_{c}$, we can solve $dp(\theta)/dt=0$ to obtain the steady-state
distribution. This distribution $p(\theta)$ can be viewed as a population
average, or the long-time average of one individual. It allows us
to compute the mean climbing speed, $v/v_{0}=\left\langle \cos\theta\right\rangle _{\theta}=\int d\theta\:p(\theta)\cos\theta$.

The information rate we use is mutual information between the heading
$\theta$ and the controlled update $d\theta_{c}$, per time $dt$.
For tumbling alone, this is 
\begin{equation}
i=I(\Theta;d\Theta_{c})/dt=\blue{\left\langle \lambda(\theta)\log\frac{\lambda(\theta)}{\left\langle \lambda(\theta')\right\rangle _{\theta'}}\right\rangle _{\negmedspace\theta}}.\label{eq:info}
\end{equation}
This form can be derived by considering the two possible actions,
$p(\text{tumble}|\theta)=\lambda(\theta)dt$ and $p(\text{run}|\theta)=1-\lambda(\theta)dt$.
The formula for continuous steering can be derived by allowing small
steps left or right of angle $\pm\alpha$ at rates $\beta_{\pm}(\theta)=D_{c}/\alpha^{2}\pm\mu(\theta)/2\alpha$.
We give more detail in appendix \ref{subsec:steering-via-alpha} (and
another approach in \ref{sec:info-derivation-anydim}), but the result
is an information rate proportional to the variance of $\mu(\theta)$:
\begin{equation}
\red{i=\left\langle \left[\mu(\theta)-\left\langle \mu(\theta')\right\rangle _{\theta'}\right]^{2}\right\rangle _{\theta}\Big/4D_{c}}.\label{eq:info-steering}
\end{equation}
The optimised solutions to (\ref{eq:objective}) always have controller
noise $D_{c}=D_{r}$, with the magnitude of steering $\mu(\theta)$
varying to produce high and low information rates. Explicit controller noise is not needed
for the tumble strategy, as the intrinsic noise of the Poisson process
 plays the same role.

Using equations (\ref{eq:dtheta})--(\ref{eq:info-steering}) in
objective (\ref{eq:objective}), we now have all the pieces to find
optimal tumbling and steering strategies. We show some of them in
figure \ref{fig:Numerical-2D}, and discuss their properties here before
turning to more general strategies in later sections.

For tumbling, our setup is very similar to that used by Mattingly
et. al. \citet{mattingly2021escherichia}. They drew a frontier on
the speed-information plane, at low information rates, and were able
to experimentally place \emph{E. coli} at about 70\% of optimal performance,
given the information they actually receive. In a shallow gradient,
these bacteria receive around 1\% of a bit per run. In this regime
the tumble rate is only slightly modulated, as shown in figure \refsub{fig:Numerical-2D}{A}:
\[
\blue{\lambda(\theta)=D_{r}\left[1-\sqrt{4i/D_{r}}\cos\theta+\mathcal{O}(i/D_{r})\right]}.
\]
The resulting performance is $v/v_{0}\approx\sqrt{i/4D_{r}}$ for
$i\ll D_{r}$, which is a dashed line in figure \ref{fig:Performance-2D}.
Here we assume that the turns are instantaneous, although real \emph{E.
coli} spend perhaps 10\% of their time tumbling \citet{waite2016nongenetic}.
In appendix \ref{sec:time-penalty} we derive the effect of time penalty
$\tau$ per tumble on the solution, 
and show that a large penalty $\tau > 0.69/D_r$ is needed for the rank order of strategies to change.

In the opposite limit, when information is plentiful, our problem
approaches the one studied by Strong et. al. \citet{strong1998adaptation}.
They asked what strategy gives the fastest climbing, with a finite
time penalty $\tau$ per tumble, but no cost of information. They
found a deterministic strategy: the agent should always tumble when
$\left|\theta\right|>\theta_{\mathrm{thresh}}$. To achieve high speed
when tumbles are quick, $\tau\ll1/D_{r}$, the threshold is low $\left|\theta_{\mathrm{thresh}}\right|\ll1$,
describing a behaviour in which the agent tumbles many times in succession
until landing on a nearly-uphill heading. Similar behaviour is seen
in the solution to our problem with $\tau=0$ but a large finite information
rate, $i\gg D_{r}$. We find that the following strategy, shown in
figure \refsub{fig:Numerical-2D}A, is a reasonable approximation: 
\begin{equation}
\blue{\lambda_{\mathrm{strong}}(\theta)=\begin{cases}
0 & \left|\theta\right|<\theta_{\mathrm{thresh}}\\
\lambda_{\mathrm{max}} & \text{else.}
\end{cases}}\label{eq:lambda-strong}
\end{equation}
Numerically, we can trace out the entire range between the low- and
high-information limits, giving the blue points on figure \ref{fig:Performance-2D}.

Turning now to study continuous changes of heading, we optimise both
the steering rate $\mu(\theta)$ and the controller noise $D_{c}$.
The solution always has $D_{c}=D_{r}$, and it's possible to solve
for $\mu(\theta)$ exactly in terms of Mathieu special functions. We give
details in appendix \ref{subsec:steering_solution}, but the leading
term at low information rate is: 
\[
\red{\mu(\theta)=-\sqrt{8iD_{r}}\sin\theta+\mathcal{O}(i/D_{r})}.
\]
This limit $i\ll D_{r}$ has performance $v/v_{0}\approx\sqrt{i/2D_{r}}$,
and we plot the full curve on figure \ref{fig:Performance-2D}. In
the high-information limit the heading is tightly constrained, $\left|\theta\right|\ll1$,
producing a simple linear control problem.

We see that the steering strategy has uniformly higher performance
than tumbling. But an important difference is that it is exploiting
the sign of $\theta$, in a way that tumbling could not. The optimal
tumble rate is an even function, $\lambda(\theta)=\lambda(-\theta)$,
but the optimal steering rate is odd, $\mu(\theta) = -\mu(-\theta)$.
A strategy in which the agent
is always able to steer towards the correct direction may be relevant
for a ship with a poor compass, or an organism large enough to sense
the local gradient vector. But if the agent can only observe its rate
of up-gradient motion --- as for instance a ship measuring on the
depth of water, or a bacterium sensing the rate of change of concentration
--- then it must be ignorant of the sign, and hence $\mu(\theta)$
must be even.

We can impose this symmetry constraint $\mu(\theta)=\mu(-\theta)$
when finding solutions. Without access to the sign of $\theta$, the
performance of steering is worse than tumbling, 
showing that discrete strategies can out-perform continuous adjustments.
In figure
\ref{fig:Performance-2D}, the low-information limit is $v/v_{0}\approx\sqrt{i/8D_{r}}$
for $i\ll D_{r}$. 
The strategy in this limit is to always turn in one
direction, but slightly bias the rate: 
\[
\orange{\mu(\left|\theta\right|)=\pm2D_{r}\Big[1-\sqrt{2i/D_{r}}\cos\theta+\mathcal{O}(i/D_{r})\Big]}.
\]
At high information rates, the strategy is more interesting, see figure
\refsub{fig:Numerical-2D}D. The optimal $\mu(\left|\theta\right|)$
changes sign to create a pair of fixed points at $\pm\theta_{\star}$,
and the agent spends most of its time at whichever one is stable.
For each such strategy, turning the other way %
gives equivalent performance, but moves across the gradient at speed
$v_{0}\left\langle \sin\theta\right\rangle _{\theta}$ with the opposite
sign.

\section*{Transitions between turn strategies}

In the absence of directional information, we have seen that tumbles,
i.e. turns by a randomly chosen angle $\Delta\theta\sim U(-\pi,\pi)$,
perform better than continuous steering. It is possible to do better
by controlling the turn angle, and the best strategy
at low information rates is in fact to \emph{reverse} direction, $\Delta\theta=\pi$.
The Fokker-Planck equation for this strategy is 
\[
\frac{dp(\theta)}{dt}=\green{-\lambda_{\mathrm{rev}}(\theta)p(\theta)+\lambda_{\mathrm{rev}}(\theta+\pi)p(\theta+\pi)}+D_{r}\frac{d^{2}p(\theta)}{d\theta^{2}}
\]
and the information rate is still (\ref{eq:info}). The optimal rate
of reversal at low information rates is (appendix \ref{sec:Parametric_ansatz_reverse},
figure \refsub{fig:Numerical-2D}B)
\[
\green{\lambda_{\mathrm{rev}}(\theta)=D_{r}/2\left[1-\sqrt{8i/D_{r}}\cos\theta+\mathcal{O}(i/D_{r})\right]}.
\]
This strategy uses half as much information for the same speed as
tumbling: $v/v_{0}\approx\sqrt{i/2D_{r}}$ for $i\ll D_{r}$. 
Curiously, in this limit reverse achieves the same performance as the steering solution $\mu(\theta)$.
But at high information rates, the best that it can do is to place all
probability density within $\left|\theta\right|<\pi/2$ (shown in
figure \refsub{fig:Numerical-2D}B), leading to $\left\langle \cos\theta\right\rangle _{\theta}=2/\pi\approx0.64$.
Hence figure \ref{fig:Performance-2D} shows a crossover: a transition
between reverse and tumble being the best strategy, among the three
not using the sign of $\theta$. %

In  appendix \ref{sec:analytic-discrete} we consider a \emph{flick}
strategy which uses right-angle turns, $\Delta\theta=\pm\pi/2$. This
strategy beats both tumble and reverse in a medium-information regime
(around $i/D_{r}\approx10$), but cannot exceed $\left\langle \cos\theta\right\rangle _{\theta}=\sqrt{8}/\pi\approx0.9$.
Thus we see two transitions between simple discrete strategies as
the amount of information available increases.

\section*{Discrete angles from arbitrary turns}

\label{sec:Discrete-angles}

Instead of a fixed turn angle $\Delta\theta$ or a fixed distribution,
we now look at the general case where $\lambda(\Delta\theta,\theta)$
is the rate of initiating turns by $\Delta\theta$ from heading $\theta$.
All of the strategies already considered can be written as such a
rate: 
\[
\lambda(\Delta\theta,\theta)=\begin{cases}
\blue{\lambda(\theta)/2\pi} & \text{tumble}\\
\green{\delta(\Delta\theta-\pi)\,\lambda_{\mathrm{rev}}(\theta)} & \text{reverse}\\
\red{\delta(\Delta\theta-\alpha)\,\beta_{+}(\theta)+\delta(\Delta\theta+\alpha)\,\beta_{-}(\theta)}\negthickspace\negthickspace\negthickspace & \text{steering.}
\end{cases}
\]
Here we again approximate  steering as turns by small angles $\Delta\theta=\pm\alpha$
with rates $\beta_{\pm}(\theta)=D_{c}/\alpha^{2}\pm\mu(\theta)/2\alpha$,
as shown in figure \refsub{fig:Numerical-2D}C. For the general case
the Fokker-Planck equation  reads
\[
\frac{dp(\theta)}{dt}=\int\negmedspace d\phi\Big[\lambda(\theta-\phi,\phi)p(\phi)-\lambda(\phi,\theta)p(\theta)\Big]+D_{r}\frac{d^{2}p(\theta)}{d\theta^{2}}
\]
and the information rate is 
\begin{equation}
i=\iint\negmedspace d\Delta\theta\:d\theta\:\lambda(\Delta\theta,\theta)p(\theta)\log\frac{\lambda(\Delta\theta,\theta)}{\bar{\lambda}(\Delta\theta)}\label{eq:info-anyjump}
\end{equation}
where we define $\bar{\lambda}(\Delta\theta)=\int d\theta'\:\lambda(\Delta\theta,\theta')p(\theta')$.

Solving numerically for the optimal $\lambda(\Delta\theta,\theta)$,
we recover solutions equivalent to steering, which are shown as red
points on figure \ref{fig:Performance-2D}. The close agreement on
performance is numerical evidence that steering via $\mu(\theta)$
is truly optimal; we hope to present analytic results on this question
soon \citet{leighton2026information}. These unconstrained solutions
obey $\lambda(\Delta\theta,\theta)=\lambda(-\Delta\theta,-\theta)$,
which aligns with an odd steering rate, $\mu(-\theta)=-\mu(\theta)$; see appendix \ref{subsec:steering-via-alpha} for examples.

    \begin{figure*}
    \centering \includegraphics[width=0.95\textwidth]{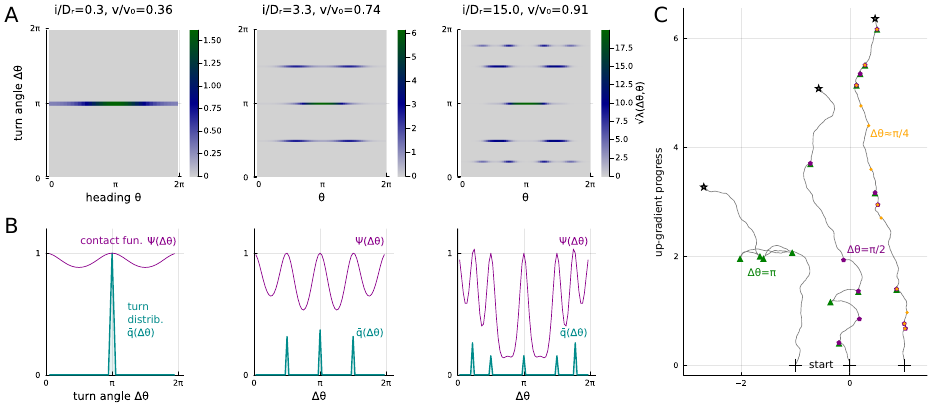}
    
    \caption{Discrete optimal strategies for three information rates. (A) We impose
    that rate $\lambda(\Delta\theta,\theta)$ is even in $\Delta\theta$,
    which implies that it is even in $\theta$, i.e. the strategy ignores the sign of $\theta$. At
    low information rate (left) we recover the reverse strategy, $\Delta\theta=\pi$,
    but with increasing information it bifurcates to use three angles
    (centre), and then five (right). The two new strategies shown are
    faster than both reverse and tumbling: see the black stars on figure
    \ref{fig:Performance-2D}. (B) Below each rate, we plot the contact
    function $\Psi(\Delta\theta)$ and a mean distribution $\bar{q}(\Delta\theta)\propto\bar{\lambda}(\Delta\theta)$.
    The rate $\lambda(\Delta\theta,\theta)$ is nonzero precisely where
    $\Psi(\Delta\theta)=1$. Notice that $\Psi(0)=1$, but we do not allow
    turns of $\Delta\theta=0$.
    (C) Sample trajectories for the same three strategies, all for times $0\leq t \leq 7$ (units $D_r=v_0=1$). Notice that in the second case, a flick turn by $\Delta\theta=\pm\pi/2$ (purple) is often followed by a reversal, $\Delta\theta=\pi$ (green), when the flick happened to pick the wrong direction.
    This strategy isn't quite run-reverse-flick \cite{xie2011bacterial}, as one flick can follow another flick, each turn is independently chosen.
    }
    \label{fig:Any-jump}
    \end{figure*}

However, if we look for strategies not using the sign
of $\theta$, then something more interesting happens, shown in figure
\ref{fig:Any-jump}. At low information rates, we recover the reverse
strategy, which is the true optimum. But as more information becomes
available, the solution bifurcates to use two additional angles, $\Delta\theta\approx\pm\pi/2$.
With more information, it bifurcates again to use five angles, and
so on, but the distribution of turn angles remains discrete.

To see why this discreteness of $\Delta\theta$ emerges, we now formulate an augmented
problem in which the Fokker-Planck equation $dp(\theta)/dt=0$ is
imposed as a constraint, along with the symmetry $\lambda(\Delta\theta,\theta)=\lambda(-\Delta\theta,\theta)$,
and the normalisation of $p(\theta)$. These are enforced by Lagrange
multipliers, and thus %
we maximise the following $\mathcal{L}$ with respect to $p,\lambda,\chi,\varphi,\xi$:
\begin{align*}
\mathcal{L} & =\left\langle \cos\theta\right\rangle _{\theta}-\gamma i+\int\negmedspace d\theta\:\chi(\theta)\frac{dp(\theta)}{dt}+\varphi\Big[1-\int\negmedspace d\theta\,p(\theta)\Big]\\
 & +\iint\negmedspace d\Delta\theta\,d\theta\,\xi(\Delta\theta,\theta)\Big[\lambda(\Delta\theta,\theta)-\lambda(-\Delta\theta,\theta)\Big].
\end{align*}
In addition to the three equality constraints, there are still two inequality constraints: $p(\theta)\geq0$ and
$\lambda(\Delta\theta,\theta)\geq0$. This last constraint plays a
crucial role, as the equation of motion $\partial\mathcal{L}/\partial\lambda(\Delta\theta,\theta)=0$
need only be satisfied where the constraint is slack, $\lambda(\Delta\theta,\theta)>0$.
After some algebra, this equation of motion reads 
\[
\lambda(\Delta\theta,\theta)/\bar{\lambda}(\Delta\theta)=e^{\left[\chi(\theta+\Delta\theta)+\chi(\theta-\Delta\theta)-2\chi(\theta)\right]/2\gamma}.
\]
Taking the expectation value with $p(\theta)$ on both sides leads
to $\Psi(\Delta\theta)=1$, where we define 
\begin{equation}
\Psi(\Delta\theta)=\int\negmedspace d\theta\:p(\theta)\:e^{\left[\chi(\theta+\Delta\theta)+\chi(\theta-\Delta\theta)-2\chi(\theta)\right]/2\gamma}.\label{eq:contact-fun-maintext}
\end{equation}
This contact function is analytic, and clearly has $\Psi(0)=1$. Thus
it must either be constant, or else have $\Psi(\Delta\theta)=1$ at a set
of isolated points. In appendix \ref{sec:discrete-target} we show
that $\Psi''(0)=-i/D_{r}<0$, ruling out the constant case. Hence
$\lambda(\Delta\theta,\theta)$ is zero except at a discrete set of
angles $\Delta\theta$. Figure \ref{fig:Any-jump} plots the contact
function alongside the numerical solutions $\lambda(\Delta\theta,\theta)$
used to find it, for three information rates.

A similar construction without the symmetry constraint 
leads to the contact function being $\Psi(\Delta\theta)=\int d\theta\:p(\theta)\:e^{\left[\chi(\theta+\Delta\theta)-\chi(\theta)\right]/\gamma}$.
In the appendix we show that, when evaluated on steering solutions,
this is 1 everywhere. The symmetry constraint may be thought of as
changing a circular parameter space $0\leq\Delta\theta<2\pi$ into a
line $0\leq\Delta\theta\leq\pi$ with reflecting boundary conditions. Such
fold lines, and other explicit boundaries of parameter space, played
a crucial role in the emergence of discreteness in our work on Bayesian
priors \citet{mattingly2018maximizing}.

Related arguments for discreteness from analyticity have been made
in channel capacity problems, see for instance %
\citet{smith1971information,hillar2017neural,mattingly2018maximizing},
and there are some results about the number of points in the support
of the solution \citet{abbott2019scaling,wang2026improved}. The
maximisation problem we study here is more complicated, and is most
similar to that of \citet{jung2019discrete}.

\section*{Navigation in three dimensions}
\label{sec:Three-dimensions}

While some organisms do navigate on a two-dimensional plane, many
swim freely in three dimensions. Here we show that the qualitative
results above survive, and some are strengthened.

    \begin{figure}
    \includegraphics[width=1\columnwidth]{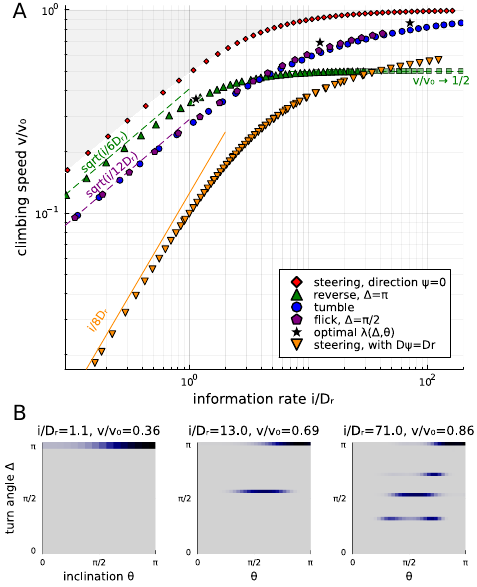}
    
    \caption{Navigation strategies in three dimensions. (A) Performance of tumble and reverse
    are qualitatively similar to figure \ref{fig:Performance-2D}, with
    different prefactors in the $v\sim\sqrt{i}$ scaling at low information
    rates. Flick produces similar performance to tumble. The red steering
    points are a strategy which controls only the inclination $\theta$,
    turning always towards the pole at $\theta=0$. The orange points
    are a steering strategy in which the agent suffers diffusion $D_{\psi}$
    in its roll angle $\psi$ (in addition to diffusion of its heading,
    $D_{r}$), and this is much worse than not knowing the sign of $\theta$
    in two dimensions, leading to performance $v\sim i$ at low information
    rates. (B) Optimal rates $\lambda(\Delta,\theta)$ for a strategy
    allowed any turn angle $\Delta$. The roll angle $\psi$ is uncontrolled,
    hence $\Delta=\pi$ (flick) represents a turn to anywhere on the circle
    perpendicular to the current heading. As in figure \ref{fig:Any-jump}
    we see a progression from reverse to reverse \& flick to something
    more complicated, but always using a discrete set of angles $\Delta$. }
    \label{fig:3D-maintext}
    \end{figure}
    
The tumble and reverse strategies have obvious generalisations to
three dimensions, and figure \refsub{fig:3D-maintext}A shows their performance.
At low information rates reverse is again faster, with $v/v_{0}\approx\sqrt{i/6D_{r}}$,
and again becomes slower than tumbling at high information rates,
now $v/v_{0}<1/2$.

Steering is more complicated, as the direction in which the agent
turns is now a vector, tangent to the sphere of possible headings.
The equivalent of making use of the sign of $\theta$ is now steering
always towards the pole, $\theta=0$. If this is possible, then the
cost of controlling $\theta$ alone is low, producing high performance:
the red points on figure \refsub{fig:3D-maintext}A. However, the agent
can also roll around its heading direction, which is described by
an additional angle $\psi$. The equivalent of not knowing the sign
of $\theta$ in two-dimensional steering is not knowing this roll
angle, and this greatly degrades the value of steering. If we assume
the roll angle suffers from diffusion with noise $D_{\psi}$, and
take $D_{\psi}=D_{r}$ then what we see numerically is, very approximately,
\[
\orange{v/v_{0}\approx i/8D_{r},\qquad i\ll D_{r}}.
\]
Analytically, we show in appendix \ref{sec:roll-steer-3D} that $v\sim i^{s}$ with slope $s\geq1$.
This is a much steeper decline at low information rate than the discrete
strategies, which all scale as $v\sim\sqrt{i}$.

The flick strategy, making right-angle turns, performs much like tumble.
One difference from two dimensions is that it now has limit $v/v_{0}\to1$
at high information rates, because a succession of right-angle turns
in three dimensions can bring the heading arbitrarily close to the
pole. Here too we assume the roll angle is not controlled, hence each
turn by $\Delta=\pi/2$ places the heading new anywhere on a circle
perpendicular to the original heading.

We can similarly study the general case where $\lambda(\Delta,\theta)$
is the rate at which turns are initiated, placing the new heading
somewhere on a circle angle $\Delta$ away from the old heading vector.
Figure \refsub{fig:3D-maintext}B shows what we see. At low information
rates, all weight is on $\Delta=\pi$, the reverse strategy. With
more information, turns of $\Delta\approx\pi/2$ appear too, and there
are further bifurcations to use additional angles. But the distribution
of turn angles remains discrete, much like in two dimensions (figure
\ref{fig:Any-jump}).

\section*{Conclusion}

In this navigation problem we see two kinds of discreteness emerge.
First, among the strategies not using directional information, sudden
actions like tumble and reverse perform better than smooth steering
(figures \ref{fig:Performance-2D}, \refsub{fig:3D-maintext}A). Second,
if arbitrary turn angles are allowed, the optimal strategy uses a
discrete set of angles (figures \ref{fig:Any-jump}, \refsub{fig:3D-maintext}B).
This has parallels to the idea of rational inattention \citet{sims2006rational,jung2019discrete},
as well as earlier work in neural coding \citet{nikitin2009neural,kastner2015critical,sharpee2017optimizing,shao2023efficient}
and elsewhere in biology \citet{tkacik2008information,witteveen2025optimizing}.

While our navigation model is far too simple to quantitatively match
data from any real organism, we can ask whether its qualitative features
show up in the wild. Certainly many micro-organisms use strategies
which alternate straight runs with sudden turns \citet{grognot2021more,grognot2021vibrio,grognot2023physiological}.
The transition from reverse to tumble being optimal (figures \ref{fig:Performance-2D},
\refsub{fig:3D-maintext}A) could map to different bacteria which have
evolved in different circumstances, or to adaptive behaviour by an
individual. For example, \emph{V. cholerae} performs a run-reverse-flick
strategy \citet{xie2011bacterial} when swimming fast, but only run-reverse
when slow \citet{son2016speeddependent}. Intriguingly, the worm \emph{C.
elegans} uses a strategy qualitatively similar to our even steering
$\mu(\left|\theta\right|)$ when certain interneurons associated with
navigation are disrupted \citet{chen2025navigation}. For our second
kind of discreteness, evidence suggests that flies adopt discrete
turn saccades with stereotyped angles \citet{demir2020walking}, and
it would be interesting to explore how this phenomenon varies with
the noise level. 

For our stated problem, the optimal solution is to steer in the direction
of $\theta=0$, the red line in Figure \ref{fig:Performance-2D}.
This could be interpreted as evidence for the high value of directional
information. %
But it's not entirely clear that we should directly compare the information
rate of such steering strategies to the others. If a limited information
rate is a proxy for sensor noise, then it is difficult to imagine
a sensor giving directional and undirected information with equal
ease. For instance with a bilateral pair of sensors, the difference
will be a noisy directional measure, while the sum will be less-noisy
but non-directional. The phase diagram of optimal strategies in this
instance is unknown.

Aside from its biological relevance, our model is an unusual control
theory problem in that it permits an exact solution for the steering
rate $\mu(\theta)$ for any information rate. The well-studied regime
there is small deviations from $\theta=0$, high information rate
$i$, but our solution still holds in the opposite limit where the
agent often makes full circles (as in figure \refsub{fig:Numerical-2D}E).
We hope to present a proof that this strategy saturates a nonlinear
performance bound soon \citet{leighton2026information}.

It is a limitation of our model that we do not consider time-dependent
or memory-based strategies, which are used by many real microorganisms.
At the simplest level, bacteria don't measure $\theta$, but get a
signal proportional to $\cos\theta$ from the derivative of attractant
concentration along their path. This involves storing knowledge about
the past for a time of order $1/D_{r}$ \citet{strong1998adaptation,clark2005bacterial,celani2010bacterial}.
Many microorganisms swim in spirals in three dimensions \citet{battista2014geometrical};
this strategy could, using memory, gather information about spatial
variations perpendicular to their track. In some situations \emph{C.
elegans} performs a weathervane motion \citet{iino2009parallel},
which could similarly leverage memory to measure spatial gradients.
We plan to explore strategies which rely on memory in future work.

Finally, the mean speed up the gradient, $v/v_{0}$, isn't the only
relevant objective. If what's being sensed is some localised food
source, then there will be scenarios where the ability to loiter near
to a point is also important \citet{clark2005bacterial}, and scenarios
where all that matters is being the first to arrive. For travelling
groups of chemotactic bacteria \citet{vo2025nongenetic}, other objectives
may be important for ensuring collective navigation of an isogenetic
population. Perhaps these objectives are amenable to similar study.

More broadly, looking beyond questions of navigation, we believe that this work illustrates the utility of asking why biology chooses discrete solutions to continuous problems.
Surely there are many other cases where evolution has arrived at a discrete strategy not because it was good enough, but because it was the best. And such cases are exciting because they can offer predictions for what should be seen under other conditions.

\section*{Acknowledgements}

We thank Marianne Bauer, Kevin Chen, Isabella Graf for comments on
the draft, and Vijay Balasubramanian, Yu Fu, Bert Kappen, Derek Sherry, Sekhar Tatikonda,
Nick Weaver for discussions.

This work was supported in part by the Yale Program in Physical and
Engineering Biology (J.M.B.), Mossman and NSERC Postdoctoral Fellowships
(M.P.L.),
NIH grants R35GM158058 (T.E. and M.C.A.), R35 GM138341 (B.B.M and M.C.A) 
and a Sloane Foundation \emph{Matter to Life} grant (T.E., B.B.M.,
M.C.A.).

\section*{Data Availability}

Code which generates the figures is available here: \href{https://github.com/mcabbott/ToSteerOrNot.jl}{github.com/mcabbott/ToSteerOrNot.jl}

\bibliographystyle{my-JHEP-4.bst}
\bibliography{My-Library}

@article{abbott2019scaling,
  title = {A scaling law from discrete to continuous solutions of channel capacity problems in the low-noise limit},
  author = {Abbott, Michael C. and Machta, Benjamin B.},
  year = 2019,
  journal = {J. Stat. Phys.},
  volume = {176},
  pages = {214--227},
  issn = {1572-9613},
  doi = {10.1007/s10955-019-02296-2},
  urldate = {2022-03-23},
  langid = {english}
}

@book{arfken2005mathematical,
  title = {Mathematical methods for physicists 6th ed.},
  author = {Arfken, George B and Weber, Hans J},
  year = 2005,
  publisher = {Elsevier}
}

@article{battista2014geometrical,
  title = {Geometrical model for malaria parasite migration in structured environments},
  author = {Battista, Anna and Frischknecht, Friedrich and Schwarz, Ulrich S.},
  year = 2014,
  journal = {Phys. Rev. E},
  volume = {90},
  pages = {042720},
  issn = {1539-3755, 1550-2376},
  doi = {10.1103/PhysRevE.90.042720},
  urldate = {2025-11-24},
  copyright = {http://link.aps.org/licenses/aps-default-license},
  langid = {english}
}

@article{berg1975how,
  title = {How bacteria swim},
  author = {Berg, Howard C.},
  year = 1975,
  journal = {Sci. Am.},
  volume = {233},
  pages = {36--44},
  issn = {0036-8733},
  doi = {10.1038/scientificamerican0875-36},
  urldate = {2025-08-12},
  langid = {english}
}

@article{celani2010bacterial,
  title = {Bacterial strategies for chemotaxis response},
  author = {Celani, Antonio and Vergassola, Massimo},
  year = 2010,
  journal = {PNAS},
  volume = {107},
  pages = {1391--1396},
  issn = {0027-8424, 1091-6490},
  doi = {10.1073/pnas.0909673107},
  urldate = {2024-10-23},
  langid = {english}
}

@article{chen2025navigation,
  title = {Navigation strategies in {{Caenorhabditis}} elegans are differentially altered by learning},
  author = {Chen, Kevin S. and Sharma, Anuj K. and Pillow, Jonathan W. and Leifer, Andrew M.},
  editor = {Sengupta, Piali},
  year = 2025,
  journal = {PLoS Biol},
  volume = {23},
  pages = {e3003005},
  issn = {1545-7885},
  doi = {10.1371/journal.pbio.3003005},
  urldate = {2026-02-06},
  langid = {english}
}

@article{clark2005bacterial,
  title = {The bacterial chemotactic response reflects a compromise between transient and steady-state behavior},
  author = {Clark, Damon A. and Grant, Lars C.},
  year = 2005,
  journal = {PNAS},
  volume = {102},
  pages = {9150--9155},
  issn = {0027-8424, 1091-6490},
  doi = {10.1073/pnas.0407659102},
  urldate = {2025-07-31},
  langid = {english}
}

@article{demir2020walking,
  title = {Walking {{Drosophila}} navigate complex plumes using stochastic decisions biased by the timing of odor encounters},
  author = {Demir, Mahmut and Kadakia, Nirag and Anderson, Hope D and Clark, Damon A and Emonet, Thierry},
  year = 2020,
  journal = {eLife},
  volume = {9},
  publisher = {eLife Sciences Publications, Ltd},
  issn = {2050-084X},
  doi = {10.7554/elife.57524},
  urldate = {2025-07-10},
  copyright = {http://creativecommons.org/licenses/by/4.0/},
  langid = {english}
}

@article{grognot2021more,
  title = {More than propellers: how flagella shape bacterial motility behaviors},
  shorttitle = {More than propellers},
  author = {Grognot, Marianne and Taute, Katja M},
  year = 2021,
  journal = {Current Opinion in Microbiology},
  volume = {61},
  pages = {73--81},
  issn = {13695274},
  doi = {10.1016/j.mib.2021.02.005},
  urldate = {2025-07-02},
  langid = {english}
}

@article{grognot2021vibrio,
  title = {Vibrio cholerae motility in aquatic and mucus-mimicking environments},
  author = {Grognot, Marianne and Mittal, Anisha and Mah'moud, Mattia and Taute, Katja M.},
  editor = {Alexandre, Gladys},
  year = 2021,
  journal = {Appl. Environ. Microbiol.},
  volume = {87},
  pages = {e01293-21},
  issn = {0099-2240, 1098-5336},
  doi = {10.1128/AEM.01293-21},
  urldate = {2025-07-02},
  langid = {english}
}

@article{grognot2023physiological,
  title = {Physiological adaptation in flagellar architecture improves {{Vibrio}} alginolyticus chemotaxis in complex environments},
  author = {Grognot, Marianne and Nam, Jong Woo and Elson, Lauren E. and Taute, Katja M.},
  year = 2023,
  journal = {PNAS},
  volume = {120},
  pages = {e2301873120},
  issn = {0027-8424, 1091-6490},
  doi = {10.1073/pnas.2301873120},
  urldate = {2025-07-02},
  langid = {english}
}

@incollection{hillar2017neural,
  title = {Neural network coding of natural images with applications to pure mathematics},
  booktitle = {Contemporary {{Mathematics}}},
  author = {Hillar, Christopher and Marzen, Sarah},
  editor = {Harrington, Heather and Omar, Mohamed and Wright, Matthew},
  year = 2017,
  volume = {685},
  pages = {189--221},
  publisher = {American Mathematical                     Society},
  address = {Providence, Rhode                     Island},
  doi = {10.1090/conm/685/13814},
  urldate = {2022-06-24},
  isbn = {978-1-4704-2321-6 978-1-4704-3743-5},
  langid = {english}
}

@article{hradil2006minimum,
  title = {Minimum uncertainty measurements of angle and angular momentum},
  author = {Hradil, Z and {\v R}eh{\'a}{\v c}ek, J and Bouchal, Z and {\v C}elechovsk{\`y}, R and {S{\'a}nchez-Soto}, {\relax LL}},
  year = 2006,
  journal = {Phys. Rev. Lett.},
  volume = {97},
  eprint = {quant-ph/0605137},
  pages = {243601},
  publisher = {APS},
  doi = {10.1103/PhysRevLett.97.243601},
  archiveprefix = {arXiv}
}

@article{iino2009parallel,
  title = {Parallel use of two behavioral mechanisms for chemotaxis in {{Caenorhabditis}} elegans},
  author = {Iino, Yuichi and Yoshida, Kazushi},
  year = 2009,
  journal = {J. Neurosci.},
  volume = {29},
  pages = {5370--5380},
  issn = {0270-6474, 1529-2401},
  doi = {10.1523/JNEUROSCI.3633-08.2009},
  urldate = {2026-02-25},
  copyright = {https://creativecommons.org/licenses/by-nc-sa/4.0/},
  langid = {english}
}

@article{jung2019discrete,
  title = {Discrete actions in information-constrained decision problems},
  author = {Jung, Junehyuk and Kim, Jeong Ho (John) and Mat{\v e}jka, Filip and Sims, Christopher A},
  year = 2019,
  journal = {Rev. Econ. Stud.},
  volume = {86},
  pages = {2643--2667},
  issn = {0034-6527, 1467-937X},
  doi = {10.1093/restud/rdz011},
  urldate = {2023-07-12},
  langid = {english}
}

@article{karmakar2021state,
  title = {State of the art of bacterial chemotaxis},
  author = {Karmakar, Richa},
  year = 2021,
  journal = {J Basic Microbiol},
  volume = {61},
  pages = {366--379},
  issn = {0233-111X, 1521-4028},
  doi = {10.1002/jobm.202000661},
  urldate = {2026-06-15},
  langid = {english}
}

@article{kastner2015critical,
  title = {Critical and maximally informative encoding between neural populations in the retina},
  author = {Kastner, David B. and Baccus, Stephen A. and Sharpee, Tatyana O.},
  year = 2015,
  journal = {PNAS},
  volume = {112},
  pages = {2533--2538},
  issn = {0027-8424, 1091-6490},
  doi = {10.1073/pnas.1418092112},
  urldate = {2023-11-29},
  langid = {english}
}

@article{leighton2026information,
  title = {On the information required for feedback control},
  author = {{Leighton et. al.}, Matthew P.},
  year = 2026,
  journal = {in preparation}
}

@article{mattingly2018maximizing,
  title = {Maximizing the information learned from finite data selects a simple model},
  author = {Mattingly, Henry H. and Transtrum, Mark K. and Abbott, Michael C. and Machta, Benjamin B.},
  year = 2018,
  journal = {PNAS},
  volume = {115},
  pages = {1760--1765},
  issn = {0027-8424, 1091-6490},
  doi = {10.1073/pnas.1715306115},
  urldate = {2021-04-09},
  langid = {english}
}

@article{mattingly2021escherichia,
  title = {Escherichia coli chemotaxis is information limited},
  author = {Mattingly, Henry H. and Kamino, Keita and Machta, Benjamin B. and Emonet, Thierry},
  year = 2021,
  journal = {Nat. Phys.},
  volume = {17},
  eprint = {2102.11732},
  pages = {1426--1431},
  doi = {10.1038/s41567-021-01380-3},
  urldate = {2021-04-06}
}

@article{nikitin2009neural,
  title = {Neural population coding is optimized by discrete tuning curves},
  author = {Nikitin, Alexander P. and Stocks, Nigel G. and Morse, Robert P. and McDonnell, Mark D.},
  year = 2009,
  journal = {Phys. Rev. Lett.},
  volume = {103},
  pages = {138101},
  issn = {0031-9007, 1079-7114},
  doi = {10.1103/PhysRevLett.103.138101},
  urldate = {2025-04-01},
  copyright = {http://link.aps.org/licenses/aps-default-license},
  langid = {english}
}

@article{shao2023efficient,
  title = {Efficient population coding of sensory stimuli},
  author = {Shao, Shuai and Meister, Markus and Gjorgjieva, Julijana},
  year = 2023,
  journal = {Phys. Rev. Research},
  volume = {5},
  pages = {043205},
  issn = {2643-1564},
  doi = {10.1103/PhysRevResearch.5.043205},
  urldate = {2026-02-20},
  langid = {english}
}

@article{sharpee2017optimizing,
  title = {Optimizing neural information capacity through discretization},
  author = {Sharpee, Tatyana O.},
  year = 2017,
  journal = {Neuron},
  volume = {94},
  pages = {954--960},
  issn = {08966273},
  doi = {10.1016/j.neuron.2017.04.044},
  urldate = {2023-04-04},
  langid = {english}
}

@article{sims2006rational,
  title = {Rational inattention: {{Beyond}} the linear-quadratic case},
  shorttitle = {Rational inattention},
  author = {Sims, Christopher A},
  year = 2006,
  journal = {Am. Econ. Rev.},
  volume = {96},
  pages = {158--163},
  issn = {0002-8282},
  doi = {10.1257/000282806777212431},
  urldate = {2022-03-28},
  langid = {english}
}

@article{smith1971information,
  title = {The information capacity of amplitude- and variance-constrained sclar {{Gaussian}} channels},
  author = {Smith, Joel G.},
  year = 1971,
  journal = {Information and Control},
  volume = {18},
  pages = {203--219},
  issn = {00199958},
  doi = {10.1016/S0019-9958(71)90346-9},
  urldate = {2022-09-15},
  langid = {english}
}

@article{son2016speeddependent,
  title = {Speed-dependent chemotactic precision in marine bacteria},
  author = {Son, Kwangmin and Menolascina, Filippo and Stocker, Roman},
  year = 2016,
  journal = {PNAS},
  volume = {113},
  pages = {8624--8629},
  issn = {0027-8424, 1091-6490},
  doi = {10.1073/pnas.1602307113},
  urldate = {2025-12-19},
  langid = {english}
}

@article{strong1998adaptation,
  title = {Adaptation and optimal chemotactic strategy for {{E}}. coli},
  author = {Strong, S. P. and Freedman, B. and Bialek, William and Koberle, R.},
  year = 1998,
  journal = {Phys. Rev. E},
  volume = {57},
  pages = {4604--4617},
  issn = {1063-651X, 1095-3787},
  doi = {10.1103/PhysRevE.57.4604},
  urldate = {2024-10-23},
  copyright = {http://link.aps.org/licenses/aps-default-license},
  langid = {english}
}

@article{taylor1974reversal,
  title = {Reversal of flagellar rotation in monotrichous and peritrichous bacteria: generation of changes in direction},
  shorttitle = {Reversal of flagellar rotation in monotrichous and peritrichous bacteria},
  author = {Taylor, Barry L. and Koshland, D. E.},
  year = 1974,
  journal = {J. Bacteriol.},
  volume = {119},
  pages = {640--642},
  issn = {0021-9193, 1098-5530},
  doi = {10.1128/jb.119.2.640-642.1974},
  urldate = {2026-02-26},
  langid = {english}
}

@article{tkacik2008information,
  title = {Information flow and optimization in transcriptional regulation},
  author = {Tkacik, G. and Callan, C. G. and Bialek, W.},
  year = 2008,
  journal = {PNAS},
  volume = {105},
  eprint = {0705.0313},
  pages = {12265--12270},
  issn = {0027-8424, 1091-6490},
  doi = {10.1073/pnas.0806077105},
  urldate = {2022-02-04},
  archiveprefix = {arXiv},
  langid = {english}
}

@article{vo2025nongenetic,
  title = {Nongenetic adaptation by collective migration},
  author = {Vo, Lam and Avgidis, Fotios and Mattingly, Henry H. and Edmonds, Karah and Burger, Isabel and Balasubramanian, Ravi and Shimizu, Thomas S. and Kazmierczak, Barbara I. and Emonet, Thierry},
  year = 2025,
  journal = {PNAS},
  volume = {122},
  pages = {e2423774122},
  issn = {0027-8424, 1091-6490},
  doi = {10.1073/pnas.2423774122},
  urldate = {2025-02-28},
  langid = {english}
}

@article{vrehavcek2008experimental,
  title = {Experimental test of uncertainty relations for quantum mechanics on a circle},
  author = {{\v R}eh{\'a}{\v c}ek, J and Bouchal, Z and {\v C}elechovsk{\`y}, R and Hradil, Z and {S{\'a}nchez-Soto}, {\relax LL}},
  year = 2008,
  journal = {Phys. Rev. A},
  volume = {77},
  eprint = {0712.0230},
  pages = {032110},
  publisher = {APS},
  doi = {10.1103/PhysRevA.77.032110},
  archiveprefix = {arXiv}
}

@article{waite2016nongenetic,
  title = {Non-genetic diversity modulates population performance},
  author = {Waite, Adam James and Frankel, Nicholas W and Dufour, Yann S and Johnston, Jessica F and Long, Junjiajia and Emonet, Thierry},
  year = 2016,
  journal = {Molecular Systems Biology},
  volume = {12},
  pages = {895},
  issn = {1744-4292, 1744-4292},
  doi = {10.15252/msb.20167044},
  urldate = {2026-01-30},
  langid = {english}
}

@article{wang2026improved,
  title = {An improved lower bound on cardinality of support of the amplitude-constrained {{AWGN}} channel},
  author = {Wang, Haiyang and Barletta, Luca and Dytso, Alex},
  year = 2026,
  journal = {IEEE Trans. Inform. Th.},
  eprint = {2512.22691},
  pages = {1--1},
  issn = {0018-9448, 1557-9654},
  doi = {10.1109/TIT.2026.3697668},
  urldate = {2026-06-07},
  archiveprefix = {arXiv},
  copyright = {https://ieeexplore.ieee.org/Xplorehelp/downloads/license-information/IEEE.html}
}

@article{witteveen2025optimizing,
  title = {Optimizing information transmission in optogenetic {{Wnt}} signaling},
  author = {Witteveen, Olivier and Rosen, Samuel J. and Lach, Ryan S. and Wilson, Maxwell Z. and Bauer, Marianne},
  year = 2026,
  journal = {Phys. Rev. Research},
  eprint = {2506.22633},
  doi = {10.1103/f7qj-f7qy},
  urldate = {2025-07-07},
  archiveprefix = {arXiv}
}

@article{xie2011bacterial,
  title = {Bacterial flagellum as a propeller and as a rudder for efficient chemotaxis},
  author = {Xie, Li and Altindal, Tuba and Chattopadhyay, Suddhashil and Wu, Xiao-Lun},
  year = 2011,
  journal = {PNAS},
  volume = {108},
  pages = {2246--2251},
  issn = {0027-8424, 1091-6490},
  doi = {10.1073/pnas.1011953108},
  urldate = {2025-07-02},
  langid = {english}
}

\onecolumngrid

\appendix

\renewcommand{\thefigure}{S\arabic{figure}}
\setcounter{figure}{0}

\renewcommand{\thesection}{\Alph{section}}
\renewcommand{\thesubsection}{\negthinspace.\arabic{subsection}}

\titleformat{\section}{\raggedright\bfseries\large}{Appendix \Alph{section}.}{1em}{}
\titlespacing\section{0pt}{10pt plus 2pt minus 1pt}{5pt} %

\titleformat{\subsection}{\raggedright\bfseries}{\Alph{section}\thinspace\thesubsection}{1em}{}
\titleformat{\subsubsection}{\raggedright\bfseries\itshape}{\thesubsubsection}{1em}{}

\newpage
\section{Additional figures}
\label{sec:more-figures}

\begin{figure}
\includegraphics[width=0.7\columnwidth]{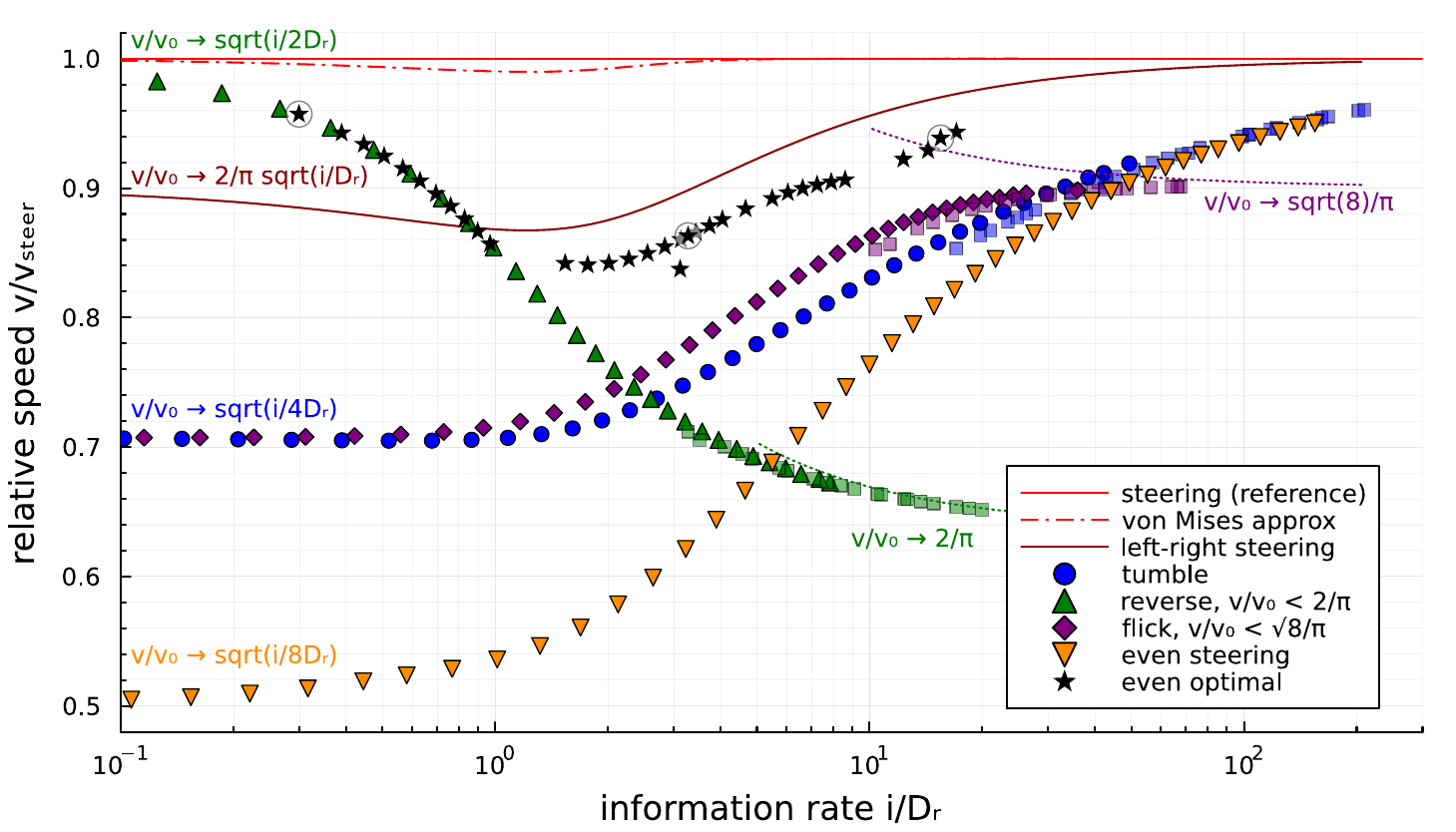}

\caption{Performance of strategies for two-dimensional navigation, relative to steering. Shows most of the data in figure \ref{fig:Performance-2D}, but adds a lines for the exact left-right steering solution (dark red, section \ref{sec:signonly}), and numerical points for the flick solution (purple) and more optimal symmetric solutions $\lambda(\lvert\Delta\theta\rvert,\theta)$ (black stars). 
The three lines are strategies exploiting the sign of $\theta$, with odd $\mu(\theta)$.
If we ignore the black stars, then notice that three different even solutions are optimal in turn -- reverse at low information rate, then flick, then tumble. As before, square plot points indicate the use of the ansatz $\lambda_\mathrm{strong}(\theta)$, while others solve for the whole function. 
\label{fig:relative-performance}
} 
\end{figure}

Figure \ref{fig:Performance-2D} in the main text compares the performance of four strategies, and observes that steering (the red line) is fastest. Figure \ref{fig:relative-performance} shows the same data plotted relative to the steering solution, and adds several more strategies.

\begin{figure}
\centering \includegraphics[width=1.0\columnwidth]{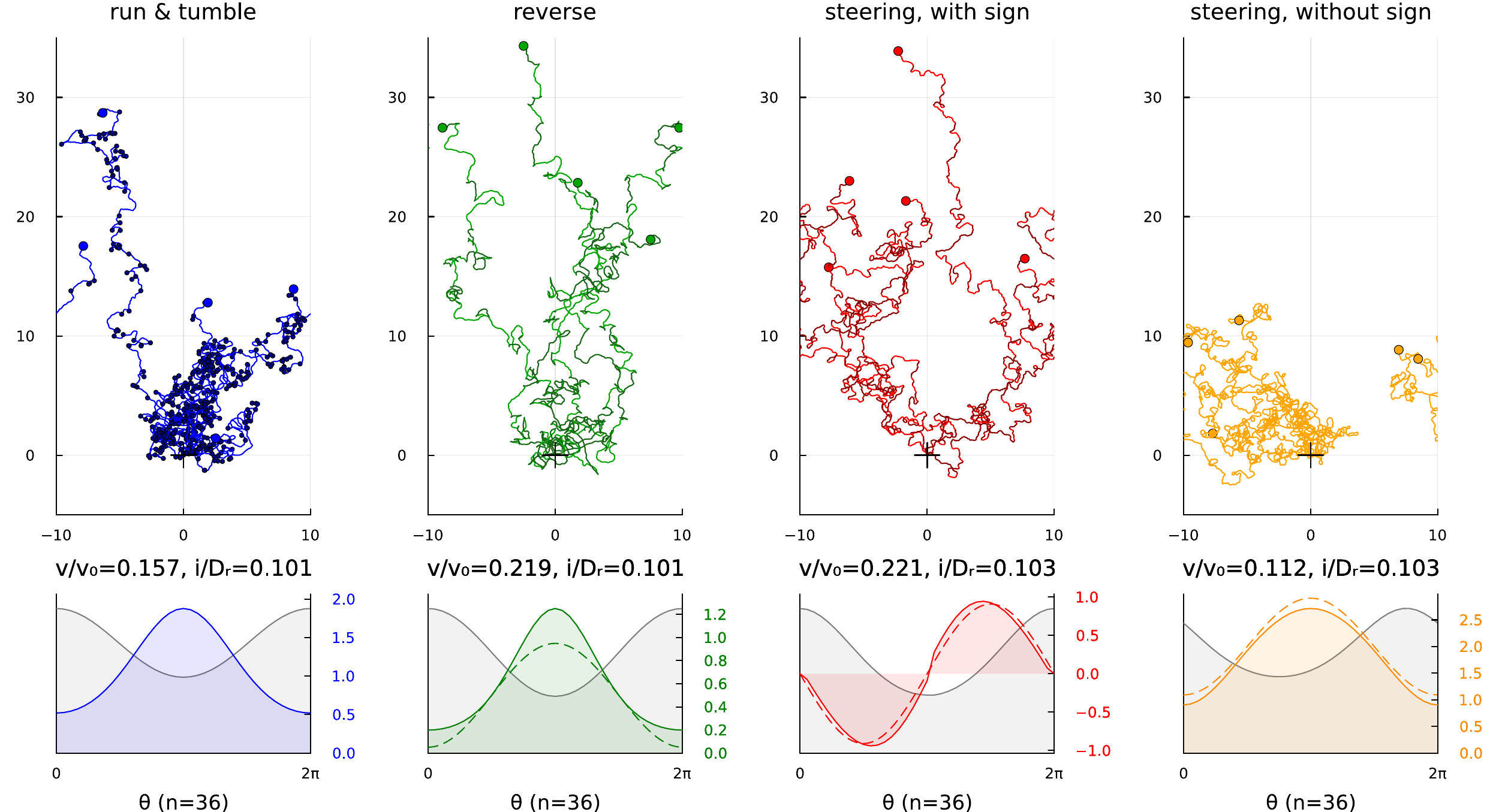}\caption{Sample trajectories for four strategies, all with the same information rate $i/D_{r}\approx0.1$. Similar to as figure \refsub{fig:Numerical-2D}E, except showing five examples of each strategy. Each reversal changes the line colour between dark and light green. Steering changes between light and dark red according to the sign of $\mu(\theta)$. At low information rates, even steering $\mu(\lvert\theta\rvert)$ has $\langle\cos\theta\rangle_\theta \approx \pm \langle\sin\theta\rangle_\theta$, so the agent goes sideways as much as up the gradient. All plots are for time $0<t<100$, and wrapped to $-10<x<10$, in units $D_{r}=v_0=1$.
}\label{fig:more-trajectories}
\end{figure}

\begin{figure}
\centering \includegraphics[width=1.0\columnwidth]{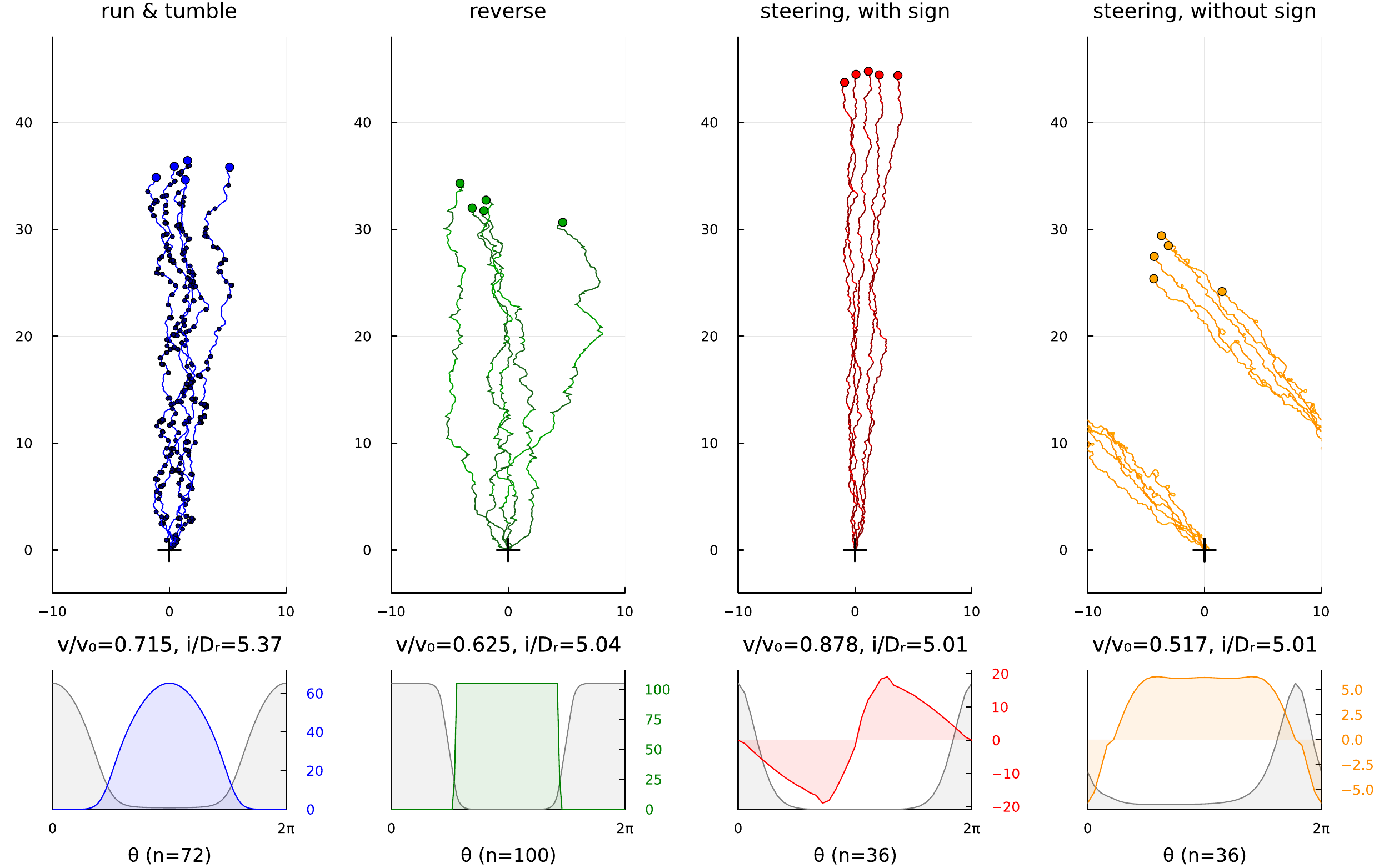}\caption{Sample trajectories for four strategies, with a higher information rate $i/D_{r}\approx5$. All plots for time $0<t<50$. Compared to figure \ref{fig:more-trajectories}, these are half the duration, with 50x the information rate. 
}\label{fig:more-trajectories-fast}
\end{figure}

Figures \ref{fig:more-trajectories} and \ref{fig:more-trajectories-fast} show some more numerical trajectories, first with the same parameters as figure \refsub{fig:Numerical-2D}E in the main text (all $i/D_r\approx 0.1$), and then with much higher information rates ($i/D_r\approx 5$). Observe particularly the steering solutions without the sign (rightmost panels) which make progress diagonally, and make occasional full turns. We make a sign choice in plotting these, as there is always an equivalent strategy, $-\mu(\theta)$, which goes diagonally the other way.

\newcommand{\wasvd}{v}
\newcommand{\waslambdac}{\lambda_c}

\section{Analytic results for factorised discrete strategies}
\label{sec:analytic-discrete}

Here we derive some results for solutions whose jump rate is of the form $\lambda(\Delta\theta,\theta) = \lambda(\theta) q(\Delta\theta)$, where $q$ is a probability distribution over target angles. We also assume that the target distribution is symmetric, such that $q(\Delta\theta)=q(-\Delta\theta)$. The tumble, reverse, and flick strategies are in this class.%

At low information rates, we study small deviations from constant $\lambda(\theta)$, and are able to derive the $v\propto \sqrt{i}$ scaling law. We then show how to introduce a finite time-penalty per discrete action, and derive the effect on the $v\propto \sqrt{i}$ law.
Finally, in Section \ref{sec:Parametric_ansatz_reverse} we make an ansatz for  steady-state $p(\theta)$ to be a von Mises distribution, which works well for the reverse strategy.

For this class of strategies, the Fokker-Planck equation reads
\begin{align}\label{eq:master-factorised}
    \frac{dp(\theta)}{dt} = - \lambda(\theta) p(\theta) + \int d\Delta \theta \, \lambda(\theta - \Delta \theta) p(\theta - \Delta \theta) q(\Delta \theta) + D_r p''(\theta),
\end{align}
and the information rate is the same as \eqref{eq:info}
\begin{align}
    i = \int d\theta \, p(\theta) \lambda(\theta) \log \left( \frac{\lambda(\theta)}{\langle \lambda \rangle} \right),
\end{align}
where $\langle \lambda \rangle = \langle \lambda(\theta) \rangle_\theta = \int d\theta \, p(\theta) \lambda(\theta)$.

\subsection{Solution at low information rate} %
At low information rate, we use a perturbative ansatz. We posit that there is some parameter $\epsilon$, which vanishes as $\gamma \to \infty$, such that our strategy can be written as
\begin{align}
    \lambda(\theta) = \waslambdac (1 + \epsilon \lambda^{(1)}(\theta) + \epsilon^2 \lambda^{(2)}(\theta) + \ldots),
\end{align}
where the only dependency on $\gamma$ is through $\epsilon$ and potentially $\waslambdac$. Since we have the freedom to choose $\waslambdac$ and $\epsilon$ to set the scale, we can choose our first-order perturbation to satisfy
\begin{align}
    \int d\theta \, \lambda^{(1)}(\theta) = 0, \quad \frac{1}{2\pi} \int d\theta \, \lambda^{(1)}(\theta)^2 = 1.
\end{align}
Intuitively, we allow for a non-zero rate as $\epsilon \to 0$ since all constant strategies use no information, but also have zero drift velocity. The second constraint ensures  that the scale of the perturbation is fixed by $\epsilon$.

We also expand the stationary density in $\epsilon$:
\begin{align}
    p(\theta) = \frac{1}{2\pi}(1 + \epsilon p^{(1)}(\theta) + \epsilon^2 p^{(2)}(\theta) + \ldots).
\end{align}
Due to normalization, we must have
\begin{align}
    \int d\theta \, p^{(k)}(\theta) = 0.
\end{align}
Under these choices, the rest of the structure of the solution should arise from optimality. We allow the distribution $q(\Delta \theta)$ to be arbitrary, as long as it is symmetric.

Using our ansatz in the master equation and collecting the first-order terms in $\epsilon$, we obtain
\begin{align}
    0 = \frac{D_r}{2\pi} \partial_\theta^2 p^{(1)}(\theta)- \frac{\waslambdac}{2\pi}[p^{(1)}(\theta) + \lambda^{(1)}(\theta)] + \frac{\waslambdac}{2\pi} \int d\theta' \, [p^{(1)}(\theta') + \lambda^{(1)}(\theta')] q(\theta - \theta').
\end{align}
We now transform this equation to obtain its Fourier coefficients, using the following convention:
\begin{align}
    g_n = \frac{1}{2\pi} \int d\theta \, e^{-in\theta} g(\theta).
\end{align}
With this, the Fourier modes of our first-order corrections satisfy
\begin{align}
    0 = -D_r n^2 p^{(1)}_n - \waslambdac[p^{(1)}_n + \lambda^{(1)}_n] + 2 \pi \waslambdac [p^{(1)}_{n} + \lambda^{(1)}_{n}] q_n.
\end{align}
Solving for $p^{(1)}_n$ we obtain
\begin{align}
    p^{(1)}_n = - \left( \frac{\waslambdac [1 - 2\pi q_n]/D_r}{n^2 + \waslambdac [1 - 2\pi q_n]/D_r} \right) \lambda^{(1)}_n.
\end{align}
Therefore, our drift velocity can be written explicitly in terms of our strategy $(\lambda, q)$:
\begin{align}
    \langle \cos(\theta) \rangle_\theta = \epsilon \frac{p^{(1)}_1 + p^{(1)}_{-1}}{2} + {\cal O}(\epsilon^2) = -\frac{\epsilon}{2} \left( \frac{\waslambdac[1 - 2\pi q_1]/D_r}{1 + \waslambdac[1 - 2\pi q_1]/D_r} \right) [\lambda^{(1)}_1 + \lambda^{(1)}_{-1}] + {\cal O}(\epsilon^2),
\end{align}
where we used the symmetry of $q$ to conclude that $q_1 = q_{-1} \in \mathbb{R}$.

Just like we can do a perturbative expansion of the drift velocity, we can do a perturbative expansion of the information rate. The leading order of this expansion will be $\epsilon^2$, so we want to keep all terms up to this order. First, we compute the average
\begin{align}
    \langle \lambda \rangle = \frac{\waslambdac}{2\pi} \int d\theta \, [1 + \epsilon p^{(1)}(\theta) + \epsilon^2 p^{(2)}(\theta) + \ldots] [1 + \epsilon \lambda^{(1)}(\theta) + \epsilon^2 \lambda^{(2)}(\theta) + \ldots].
\end{align}
We obtain
\begin{align} \label{eq:avg_lambda_expansion}
    \langle \lambda \rangle = \waslambdac\left[1 + \epsilon^2 \lambda^{(2)}_0 + \epsilon^2 \frac{1}{2\pi} \int d\theta \, p^{(1)}(\theta) \lambda^{(1)}(\theta) + {\cal O}(\epsilon^3) \right].
\end{align}
Doing a perturbative expansion of the logarithm, this implies that
\begin{align}
    \log\left( \frac{\waslambdac}{\langle \lambda \rangle} \right) = -\epsilon^2 \lambda^{(2)}_0 - \epsilon^2 \frac{1}{2\pi} \int d\theta \, p^{(1)}(\theta) \lambda^{(1)}(\theta) + {\cal O}(\epsilon^3).
\end{align}
Similarly, we have that
\begin{align}
    \log\left( \frac{\lambda(\theta)}{\waslambdac} \right) = \epsilon \lambda^{(1)}(\theta) + \epsilon^2 \lambda^{(2)}(\theta) - \frac{\epsilon^2}{2} \lambda^{(1)}(\theta)^2 + {\cal O}(\epsilon^3).
\end{align}
Grouping terms together, we obtain
\begin{align}
    \log\left( \frac{\lambda(\theta)}{\langle \lambda \rangle} \right) =  \epsilon \lambda^{(1)}(\theta) + \epsilon^2 \left[ \lambda^{(2)}(\theta) - \lambda_0^{(2)} - \frac{1}{2} \lambda^{(1)}(\theta)^2 - \frac{1}{2\pi} \int d\theta' \, p^{(1)}(\theta') \lambda^{(1)}(\theta') \right] + {\cal O}(\epsilon^3).
\end{align}
The expansion for $p(\theta) \lambda(\theta)$ is
\begin{align}
    p(\theta) \lambda(\theta) = \frac{\waslambdac}{2\pi} \left[ 1 + \epsilon (p^{(1)}(\theta) + \lambda^{(1)}(\theta)) + {\cal O}(\epsilon^2) \right].
\end{align}
These two expansions are sufficient to obtain all the terms to order $\epsilon^2$.

Let us now decompose the information rate into two terms: $i = i_1 + i_2 + {\cal O}(\epsilon^3)$, given by
\begin{align}
    i_1 &= \epsilon^2 \frac{\waslambdac}{2 \pi} \int d\theta \, [p^{(1)}(\theta) + \lambda^{(1)}(\theta)] \lambda^{(1)}(\theta), \nonumber \\
    i_2 &= \epsilon^2 \frac{\waslambdac}{2 \pi} \int d\theta \, \left[ \lambda^{(2)}(\theta) - \lambda_0^{(2)} - \frac{1}{2} \lambda^{(1)}(\theta)^2 - \frac{1}{2\pi} \int d\theta' \, p^{(1)}(\theta') \lambda^{(1)}(\theta') \right].
\end{align}
In $i_2$, we can easily see that the first two terms cancel, so we obtain
\begin{align}
    i_2 = -\epsilon^2 \frac{\waslambdac}{2\pi} \int d\theta \, \left[p^{(1)}(\theta) \lambda^{(1)}(\theta) + \frac{1}{2} \lambda^{(1)}(\theta)^2 \right].
\end{align}
Bringing these terms together, we obtain the leading-order term for the information rate:
\begin{align}
    i = \frac{\epsilon^2 \waslambdac}{4 \pi} \int d\theta \, \lambda^{(1)}(\theta)^2 + {\cal O}(\epsilon^3) = \frac{1}{2} \epsilon^2 \waslambdac + {\cal O}(\epsilon^3),
\end{align}
where the last equality comes from the choice of scale we made in our expansion.

We can now state the optimization problem for the leading-order terms in drift velocity and information rate:
\begin{align}
    \max_{\lambda^{(1)}, \waslambdac, \epsilon} \Bigg\{ \underbrace{-\frac{\epsilon}{2} \left( \frac{\waslambdac[1 - 2\pi q_1]/D_r}{1 + \waslambdac[1 - 2\pi q_1]/D_r} \right) [\lambda^{(1)}_1 + \lambda^{(1)}_{-1}]}_{\wasvd/v_0} - \gamma \underbrace{\frac{1}{2} \epsilon^2 \waslambdac}_{i} \Bigg\}.
\end{align}
Note that we allow $\epsilon$ to be chosen optimally. Our solution is consistent if the optimal epsilon vanishes as $\gamma \to \infty$. Additionally, we keep our $q$ strategy fixed for now. First, note that the information rate does not depend on the structure of the $\lambda^{(1)}$ modes. Therefore, it is optimal to just keep the $n = \pm 1$ modes and choose them to be $-1/\sqrt{2}$, such that $\lambda^{1}(\theta) = - \sqrt{2} \cos(\theta)$. The optimization problem, then, becomes
\begin{align}
    \max_{\waslambdac, \epsilon} \left\{ \frac{\epsilon}{\sqrt{2}} \left( \frac{\waslambdac[1 - 2\pi q_1]/D_r }{1 + \waslambdac[1 - 2\pi q_1]/D_r} \right) - \gamma \frac{1}{2} \epsilon^2 \waslambdac \right\}.
\end{align}
Solving for the optimal $\epsilon$ and $\waslambdac$ yields
\begin{align}
    \epsilon^* &= \frac{1 - 2 \pi q_1}{2 \sqrt{2} \gamma D_r}, \nonumber \\
    \waslambdac^* &= \frac{D_r}{1- 2 \pi q_1}.
\end{align}
This justifies our initial choice of the perturbative ansatz, since $\epsilon^*$ scales as $1/\gamma$ and $\waslambdac^*$ is independent of $\gamma$.

Under the optimal strategy, we can find a relationship between the drift velocity and the information rate, which yields our low-information Pareto frontier:
\begin{align}
    \frac{\wasvd}{v_0} = \left( \frac{i [1-2 \pi q_1]}{4 D_r} \right)^{1/2}.
\end{align}
We can see what this frontier is for some of the strategies we analyze in the main text. For run and tumble, we have $q(\Delta \theta) = 1/2\pi$, so $q_1 = 0$. Therefore, its Pareto frontier scales as
\begin{align}
    \left( \frac{\wasvd}{v_0} \right)_{\text{tumbles}} = \left( \frac{i}{4 D_r} \right)^{1/2}.
\end{align}
The target distribution for the reversing strategy is $q(\Delta \theta) = \delta(\Delta \theta - \pi)$. This means $q_1 = - 1/2\pi$, so the Pareto frontier is
\begin{align}
    \left( \frac{\wasvd}{v_0} \right)_{\text{reverse}} = \left( \frac{i}{2 D_r} \right)^{1/2}.
\end{align}
In fact, reversing achieves the highest possible drift velocity in this regime. For any symmetric strategy $q$, we have that
\begin{align}
    q_1 = \frac{1}{2\pi} \int_{0}^{2\pi} d\Delta \theta \, e^{-i \Delta \theta} q(\Delta \theta) = \frac{1}{2\pi} \int_{0}^{2\pi} d\Delta \theta \, \cos(\Delta \theta) q(\Delta \theta) = \frac{1}{2\pi} \langle \cos(\Delta \theta) \rangle_q.
\end{align}
Since $\cos(\Delta \theta) \ge -1$, we must have
\begin{align}
    q_1 \ge - \frac{1}{2\pi}.
\end{align}
Since the reversing strategy saturates this bound, and the scaling of the frontier only depends on $q_1$, we conclude that it is the optimal low-information strategy.

In addition to the information rate, we can find the leading-order behavior of the jump rate. In terms of $\lambda^*$ and $\epsilon^*$, this is
\begin{align}
    \lambda(\theta) \approx \waslambdac^* \left[ 1 - \sqrt{2} \epsilon^* \cos(\theta) \right].
\end{align}
Using our solution and writing these in terms of $i$, we obtain
\begin{align}
    \lambda(\theta) \approx \frac{D_r}{1- 2 \pi q_1} \left[ 1 - 2 \left( \frac{i}{D_r} \right)^{1/2} (1-2\pi q_1)^{1/2} \cos(\theta) \right].
\end{align}

\subsection{Time penalty for tumbles or other actions}
\label{sec:time-penalty}

So far we have considered instantaneous changes of heading, but any real organisms takes a finite amount of time to turn. For example, tumbles take up order 10\% of an \emph{E. coli}'s time. To build this into our framework, we now introduce a time penalty $\tau$ for each tumble, or other action. We write $\bar{p}_{\tau}(t)$ for the proportion of time spent tumbling, and $\bar{p}(\theta,t)$ for the rest of the time. These obey
\begin{align}
    \bar{p}_{\tau}(t)=\int_{t-\tau}^{t}dt'\int d\theta\:\lambda(\theta)\:\bar{p}(\theta,t'),\qquad\bar{p}_{\tau}(t)+\int d\theta\:\bar{p}(\theta,t)=1.
\end{align}
The Fokker-Planck equation for $d\bar{p}(\theta,t)/dt$ takes the
same form as above, except that it needs a time delay in the source
term: $\int d\Delta \theta\:\lambda(\theta-\Delta \theta)\:\bar{p}(\theta-\Delta \theta,t-\tau) q(\Delta \theta)$. This linear equation has steady-state solution $\bar{p}(\theta,t)=\eta\:p(\theta)$, where $p(\theta)$ is the normalized density solving the original ($\tau=0$) equation above. Solving, we get
\begin{align}
    \bar{p}_{\tau}(t)=1-\eta,\qquad\eta=\frac{1}{1+\tau\left\langle \lambda(\theta)\right\rangle _{\theta}},
\end{align}
where the expectation value is still with respect to $p(\theta)$, i.e. $\left\langle \lambda(\theta)\right\rangle _{\theta}=\int d\theta\:p(\theta)\lambda(\theta)$. The up-gradient relative speed averages over only the time not spent tumbling:
\begin{align}
    v/v_{0}=\int d\theta\:\bar{p}(\theta,t)\cos\theta=\eta\left\langle \cos\theta\right\rangle _{\theta}.
\end{align}
Therefore, our optimization problem now reads:
\begin{align}
    \max_{\lambda, q}\;\frac{\left\langle \cos\theta\right\rangle _{\theta}}{1+\tau\left\langle \lambda(\theta)\right\rangle _{\theta}}-\gamma\,i.
\end{align}

In the low information regime, we can use the perturbative expansions in the previous section. From Eq. \eqref{eq:avg_lambda_expansion}, we have that the corrections to $\langle \lambda \rangle$ are of second order in $\epsilon$. Therefore, following the procedure in the previous section, we arrive at the leading-order optimization problem for a fixed $q$:
\begin{align}
    \max_{\waslambdac, \epsilon} \left\{ \frac{\epsilon}{\sqrt{2} (1 + \tau \waslambdac)} \left( \frac{\waslambdac[1 - 2\pi q_1]/D_r }{1 + \waslambdac[1 - 2\pi q_1]/D_r} \right) - \gamma \frac{1}{2} \epsilon^2 \waslambdac \right\}.
\end{align}
Recall that $q_1$ is the first Fourier mode of $q(\Delta \theta)$. Solving the problem gives the frontier
\begin{align}
    \frac{\wasvd}{v_0} = \left( \frac{i [1-2 \pi q_1]}{4 D_r} \right)^{1/2} U\left( \frac{D_r \tau}{1 - 2\pi q_1} \right),
\end{align}
where $U$ is a universal function given by
\begin{align}
    U(x) \coloneqq \frac{2 \sqrt{\phi(x)}}{(1 + x \phi(x))(1 + \phi(x))}, \quad \phi(x) \coloneqq \frac{-(1+x) + \sqrt{(1+x)^2 + 12 x}}{6 x}.
\end{align}
Importantly, $U$ is decreasing and satisfies $U(0) = 1$ and $U(x \to \infty) \to 0$. For intuition, it useful to note that $U$ satisfies
\begin{align}
    \left( \frac{1}{2 + x} \right)^{1/2} \le U(x) \le \left( \frac{1}{1 + x} \right)^{1/2}.
\end{align}

We can evaluate the value of the time delay $\tau$ for which tumbling and reversing perform worse than even steering at low information. For tumbles $q_1 = 0$, so the critical $\tau$ is determined by
\begin{align}
    U(D_r \tau_{\text{tumbles}}) = \frac{1}{\sqrt{2}} \quad \implies \quad D_r \tau_{\text{tumbles}} \approx 0.694.
\end{align}
For reversing, we have $q_1 = -1/2\pi$. Therefore, the critical $\tau$ is determined by
\begin{align}
    U\left( \frac{D_r \tau_{\text{reverse}}}{2} \right) = \frac{1}{2} \quad \implies \quad D_r \tau_{\text{reverse}} \approx 4.907.
\end{align}

\subsection{Parametric ansatz for reverse strategies}\label{sec:Parametric_ansatz_reverse}
For a given strategy, we can parametrize a form of the stationary distribution that seems consistent with the numerical solutions. In particular, we will characterize the distributions in terms of a width parameter $\kappa$, and denote them with $p_\kappa(\theta)$. Our optimization problem, then, is
\begin{align}
    \max_{\kappa, \lambda} \left\{ \langle \cos(\theta) \rangle_\theta - \gamma i\right\},
\end{align}
subject to the condition that $\lambda$ generates $p_\kappa$ as the stationary distribution. Conditional on $\kappa$, the drift velocity $\langle \cos(\theta) \rangle_\theta$ is only determined by $p_\kappa$. Therefore, we can solve the optimization process in two steps. First, define
\begin{align}
    \lambda_\kappa \coloneqq \mathop{\mathrm{argmin}}_{\lambda} i . %
\end{align}
That is, first we solve for the jump rate that minimizes the information rate subject to the master-equation constraint. Then we can perform the maximization over $\kappa$ for a given $\gamma$. However, if the problem is sufficiently nice, we only need to solve the problem (min $\lambda$) and the Pareto frontier will be traced out by varying $\kappa$.

The stationarity condition for reversing takes the form
\begin{align}
    D_r p''(\theta) - p(\theta) \lambda(\theta) + p(\theta + \pi) \lambda(\theta + \pi) = 0.
\end{align}
This gives an interesting constraint on our stationary distribution:
\begin{align}
    p''(\theta) = - p''(\theta + \pi).
\end{align}
Knowing this, and looking at the numerical results, we propose the following ansatz for the stationary distribution:
\begin{align}
    p_\kappa(\theta) = \frac{1}{\pi}\Phi(\kappa \cos(\theta)),
\end{align}
where $\Phi$ is a smooth CDF, which satisfies $\Phi(-x) = 1 - \Phi(x)$. This distribution satisfies the anti-symmetry constraint and smoothly interpolates between a uniform distribution and a uniform distribution on the interval $[-\pi/2,\pi/2]$.

We want to optimize for the jump rate that minimizes the information rate. We can write the information rate in terms of the jump rate on the interval $[-\pi/2,\pi/2]$:
\begin{align}
    i = \int_{-\pi/2}^{\pi/2} d\theta \, \left[ p(\theta) \lambda(\theta) \log\left( \frac{\lambda(\theta)}{\langle \lambda \rangle} \right) + p(\theta+\pi) \lambda(\theta+\pi) \log\left( \frac{\lambda(\theta+\pi)}{\langle \lambda \rangle} \right) \right].
\end{align}
This allows us to explicitly incorporate the constraint, since $\lambda(\theta)$ and $\lambda(\theta + \pi)$ cannot be independently varied. Using our constraint, we can write this as
\begin{align}
    i = \int_{-\pi/2}^{\pi/2} d\theta \,\left[ p(\theta) \lambda(\theta) \log\left( \frac{\lambda(\theta)}{\langle \lambda \rangle} \right) + [p(\theta)\lambda(\theta) - D_r p''(\theta)] \log\left( \frac{p(\theta) \lambda(\theta) - D_r p''(\theta)}{p(\theta+\pi)\langle \lambda \rangle} \right) \right].
\end{align}
Additionally, we can write the average jumping rate as
\begin{align}
    \langle \lambda \rangle = \int_{-\pi/2}^{\pi/2} d\theta \, [p(\theta) \lambda(\theta) + p(\theta + \pi) \lambda(\theta+\pi)].
\end{align}
With the constraint, this becomes
\begin{align}
    \langle \lambda \rangle = \int_{-\pi/2}^{\pi/2} d\theta \, [2 p(\theta) \lambda(\theta) - D_r p''(\theta)].
\end{align}
Simplifying:
\begin{align}
    \langle \lambda \rangle = 2 \int_{-\pi/2}^{\pi/2} d\theta \,p(\theta) \lambda(\theta) - D_r [p'(\pi/2) - p'(-\pi/2)].
\end{align}
This gives us our first functional derivative:
\begin{align}
    \frac{\delta \langle \lambda \rangle}{\delta \lambda(\theta)} = 2 p(\theta).
\end{align}

With this setup, we can write our first-order condition as
\begin{align}
    \frac{\delta i}{\delta \lambda(\theta)} = 0.
\end{align}
Taking this variation explicitly yields
\begin{align}
    \frac{\delta i}{\delta \lambda(\theta)} =& p(\theta) \log\left( \frac{\lambda(\theta)}{\langle \lambda \rangle} \right) + p(\theta) \log\left( \frac{p(\theta)\lambda(\theta) - D_r p''(\theta)}{p(\theta+\pi)\langle \lambda \rangle} \right) \nonumber \\
    &+ 2 p(\theta)- \int_{-\pi/2}^{\pi/2} d\theta' \, \left[ \frac{p(\theta') \lambda(\theta')}{\langle \lambda \rangle} + \frac{p(\theta') \lambda(\theta') - D_r p''(\theta')}{\langle \lambda \rangle} \right] \frac{\delta \langle \lambda \rangle}{\delta \lambda(\theta)}.
\end{align}
Note that this last integral evaluates to $2 p(\theta)$. Therefore, we can simplify the FOC to
\begin{align}
    \log\left( \frac{\lambda(\theta)}{\langle \lambda \rangle} \right) + \log\left( \frac{p(\theta)\lambda(\theta) - D_r p''(\theta)}{p(\theta+\pi)\langle \lambda \rangle} \right) = 0.
\end{align}
This implies that
\begin{align}
    \lambda(\theta) \lambda(\theta + \pi) = \langle \lambda \rangle^2.
\end{align}
Note that this result holds regardless of the parametric ansatz.

This result gets us most of the way to characterizing the solution. For a given $\kappa$, let $\tilde{\lambda}_\kappa(\theta) \coloneqq \lambda_\kappa(\theta)/\langle \lambda_\kappa \rangle$. Then our optimality condition is
\begin{align}
    \tilde{\lambda}_\kappa(\theta) \tilde{\lambda}_\kappa(\theta+\pi) = 1.
\end{align}
Multiplying by $p(\theta)p(\theta+\pi)$ and using our constraint, we obtain
\begin{align}
    p_\kappa(\theta) \tilde{\lambda_\kappa}(\theta)\left[p_\kappa(\theta) \tilde{\lambda}_\kappa(\theta) - \frac{D_r}{\langle \lambda_\kappa \rangle} p_\kappa''(\theta) \right] = p_\kappa(\theta) p_\kappa(\theta+\pi).
\end{align}
Another nice property of our ansatz is that $p_\kappa(\theta + \pi) = \frac{1}{\pi} - p_\kappa(\theta)$. Therefore, we obtain
\begin{align}
    p_\kappa(\theta) \tilde{\lambda}_\kappa(\theta)\left[p_\kappa(\theta) \tilde{\lambda}_\kappa(\theta) - \frac{D_r}{\langle \lambda_\kappa \rangle} p_\kappa''(\theta) \right] = p_\kappa(\theta) \left[ \frac{1}{\pi} - p_\kappa(\theta) \right].
\end{align}
We can solve for $p_\kappa(\theta)\tilde{\lambda}_\kappa(\theta)$:
\begin{align}
    p_\kappa(\theta) \tilde{\lambda}_\kappa(\theta) = \frac{D_r}{2 \langle \lambda_\kappa \rangle} p_\kappa''(\theta) + \left( \frac{D_r^2}{4 \langle \lambda_\kappa \rangle^2}p_\kappa''(\theta)^2 + p_\kappa(\theta) \left[ \frac{1}{\pi} - p_\kappa(\theta) \right] \right)^{1/2}.
\end{align}
Additionally, note that by definition
\begin{align}
    \int d\theta \, p_\kappa(\theta) \tilde{\lambda}_\kappa(\theta)= 1.
\end{align}
Therefore, $\langle \lambda_\kappa \rangle$ must satisfy the fixed-point equation
\begin{align}
    \langle \lambda_\kappa \rangle = \int d\theta \, \left( \frac{D_r^2}{4}p_\kappa''(\theta)^2 + p_\kappa(\theta) \left[ \frac{1}{\pi} - p_\kappa(\theta) \right] \langle \lambda_\kappa \rangle^2 \right)^{1/2}.
\end{align}
Once $\langle \lambda_\kappa \rangle$ is determined, the optimal jump rate is
\begin{align}
    \lambda_\kappa(\theta) = \frac{1}{p_\kappa(\theta)} \left[ \frac{D_r}{2} p_\kappa''(\theta) + \left( \frac{D_r^2}{4}p_\kappa''(\theta)^2 + p_\kappa(\theta) \left[ \frac{1}{\pi} - p_\kappa(\theta) \right] \langle \lambda_\kappa \rangle^2 \right)^{1/2} \right].
\end{align}
For a given $\kappa$, this gives us a routine to find a point on the estimated frontier, since we can evaluate the drift velocity and information rate using $p_\kappa(\theta)$ and $\langle \lambda_\kappa \rangle$.

\newpage
\section{Analytic results for continuous steering}
\label{sec:steering}

Here we derive analytic results for the  strategies when only continuous steering $\mu(\theta)$ is used, with no discrete jumps.
First, we derive the information rate \eqref{eq:info}; see also Section \ref{sec:info-derivation-anydim} below for a derivation in $d$ dimensions.

Then we find some optimal solutions, starting with most general version where the agent can measure the full heading $\theta$, which we call signed information. We also consider a variant where the agent has access only to the sign, left or right, for which we find a much simpler analytic solution. Finally, we treat the case of steering without the sign, where we have an analytic solution only at low information rates.

\subsection{Steering and diffusion as turns $\Delta\theta = \pm \alpha$}
\label{subsec:steering-via-alpha}

We can write the controlled update for $\theta$ for steering and diffusion in short time $\Delta t$ as follows:
\begin{equation}
\Delta\theta=\mu(\theta)\Delta t+\sqrt{2D_c\,\Delta t}\,\eta,\qquad\eta\sim\mathcal{N}(0,1).
\end{equation}
Averaging over the distribution $p(\Delta\theta|\theta)$, the mean change is $\left\langle \Delta\theta\right\rangle_{\Delta\theta\sim p(\Delta\theta|\theta)} =\mu(\theta)\Delta t$, and the variance is $\left\langle (\Delta\theta-\mu(\theta)\Delta t)^{2}\right\rangle_{\Delta\theta\sim p(\Delta\theta|\theta)} =2D_c\Delta t$.
We can reproduce these with small finite jumps of $\Delta\theta=\pm \alpha$, with rates $\beta_\pm(\theta)$:
\begin{equation}
\begin{aligned}
p(\pm\alpha|\theta) & =\beta_{\pm}(\theta)\Delta t\\
p(0|\theta) & =1-\sum_{\pm}\beta_{\pm}(\theta)\Delta t.
\end{aligned}
\end{equation}
Solving, we get
\begin{equation}
\beta_{\pm}(\theta)=\frac{D_c}{\alpha^{2}}\pm\frac{\mu(\theta)}{2\alpha}+\mathcal{O}(\Delta t).
\label{eq:beta-pm}
\end{equation}
Here we must assume $D_c>\alpha\,\lvert\mu(\theta)\rvert/2>0$.

Now we can plug these probabilities into the mutual information:
\begin{equation}
\label{eq:expanded-info}
I(\Theta;\Delta\Theta)=\int d\theta\sum_{\pm}p(\pm\alpha|\theta)p(\theta)\log\frac{p(\pm\alpha|\theta)}{p(\pm\alpha)}+\mathcal{O}(\Delta t^{2})
\end{equation}
where the denominator is
\begin{equation}
p(\pm\alpha) =\int d\theta'\,p(\pm\alpha|\theta')p(\theta')
=\left[\frac{D_c}{\alpha^{2}}\pm\frac{\left\langle \mu\right\rangle }{2\alpha}\right]\Delta t+\mathcal{O}(\Delta t^{2}).
\end{equation}
Then expanding at small $\alpha$ we get 
\begin{align}
I(\Theta;\Delta\Theta) & =\int d\theta\:p(\theta)\sum_{\pm}\left[\frac{D_c}{\alpha^{2}}\pm\frac{\mu(\theta)}{2\alpha}\right]\log\frac{\frac{D_c}{\alpha^{2}}\pm\frac{\mu(\theta)}{2\alpha}}{\frac{D_c}{\alpha^{2}}\pm\frac{\left\langle \mu\right\rangle }{2\alpha}}+\ldots\\
 & =\Delta t\int d\theta\:p(\theta)\frac{\left[\mu(\theta)-\left\langle \mu\right\rangle \right]^{2}}{4D_c}+\mathcal{O}(\alpha,\Delta t^{2})
\end{align}
which is the desired formula, equation \eqref{eq:info-steering} in the main text. 
The formula, for tumbles, \eqref{eq:info}, can be derived in much the same way, taking $p(\mathrm{tumble}|\theta) = \lambda(\theta)\Delta t$ and expanding in $\Delta t$ like \eqref{eq:expanded-info}.

\begin{figure}
\includegraphics[width=0.8\columnwidth]{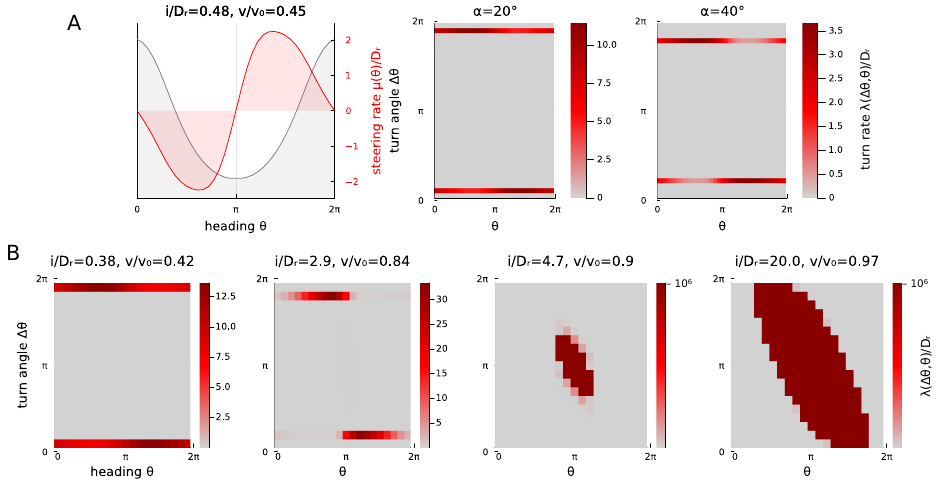}

\caption{Numerical steering solutions.
(A) Any steering rate $\mu(\theta)$ can be translated to a turn rate $\lambda(\Delta\theta,\theta)$ using angles $\Delta\theta=\pm\alpha$, and here we show the same solution translated to two different choices of $\alpha$: either small-angle turns at a high rate, or larger-angle turns at a lower rate.
(B) Some numerical solutions finding the unconstrained rate $\lambda(\Delta\theta,\theta)$ directly.
These are among the solutions plotted on as red diamonds on figure \ref{fig:Performance-2D}.
We believe that their performance has converged, but the exact rates $\lambda(\Delta\theta,\theta)$ shown here have not, and depend on details of the numerical algorithm used. This is because of the nearly flat direction in parameter space illustrated in panel A.
At low information rates, the solution shown uses the smallest turns possible, i.e. sets $\alpha$ to the discretisation scale.
But at high information rates, some solutions use larger angles, eventually pushing on the constraint  $\lambda(\Delta\theta,\theta) \leq 10^6$ imposed on the solver. 
}
\label{fig:numerical-steering}
\end{figure}

Notice that the same steering rate $\mu(\theta)$ can be encoded as turns by a larger or smaller angle $\alpha$, with larger $\alpha$ corresponding to smaller Poisson rate $\beta_\pm$ from \eqref{eq:beta-pm}. This freedom is illustrated in figure \refsub{fig:numerical-steering}A.
When solving numerically for the optimal $\lambda(\Delta\theta,\theta)$, the same freedom gives a nearly-flat direction in the parameter space. This means that the performance (speed and information, figure \ref{fig:Performance-2D}) converges, while the exact rates do not. Figure \refsub{fig:numerical-steering}B shows what we see but different solver choices will change things. ??

\subsection{Exact solution for steering $\mu(\theta)$} \label{subsec:steering_solution}
When $\mu(\theta)$ has full freedom, we expect by symmetry that there will be no net rotational flux of the heading. In this case we can integrate the Fokker-Planck equation after setting the steady-state condition $\partial_tp=0$, to obtain a Boltzmann-like solution of the form
\begin{equation}
p(\theta) = \frac{1}{Z}\exp\left(\frac{1}{D_r+D_c}\int\mu(\theta')d\theta'\right),
\end{equation}
for partition function $Z$ defined to ensure normalization. It follows then that the steering force can be written as a function of the probability distribution as
\begin{equation}
\mu(\theta) = (D_r+D_c)\partial_\theta\log p(\theta).
\end{equation}
This allows us to write the information rate as a function only of $p(\theta)$, as
\begin{subequations}
\begin{align}
i & = \frac{1}{4 D_c}\mathrm{Var}(\mu) \\
& = \frac{(D_r+D_c)^2}{4D_c}\int_0^{2\pi} p(\theta)\left[\partial_\theta\log p(\theta)\right]^2d\theta.
\end{align}
\end{subequations}
We can then write the objective function as a functional integral,
\begin{equation}
\mathcal{L} = \int_0^{2\pi}d\theta \underbrace{\left[ p(\theta)\cos\theta - \frac{\gamma(D_r+D_c)^2}{4D_c}p(\theta)\left[\partial_\theta\log p(\theta)\right]^2 + \alpha\left[p(\theta)-\frac{1}{2\pi}\right]\right]}_{f(\theta,p,p')}.
\end{equation}
Here $\alpha$ is a Lagrange multiplier ensuring normalization of $p(\theta)$. Assuming that $\mu(\theta)$ has sufficient freedom to tune $p(\theta)$ as needed for any $D_c$, we can optimize $\mathcal{L}$ with respect to $D_c$ at constant $p(\theta)$, which yields $D_c=D_r$. We then extremize $\mathcal{L}$ by deriving the Euler-Lagrange equation, obtaining the following differential equation which $p(\theta)$ must solve:
\begin{equation}
0 = \alpha + \cos\theta + \gamma D_r \left(\frac{p'(\theta)}{p(\theta)}\right)^2 + 2\gamma D_r \frac{d}{d\theta}\left( \frac{p'(\theta)}{p(\theta)} \right).
\end{equation}
This differential equation falls within the Mathieu family, and can thus be solved exactly in closed form using special functions; after enforcing normalization and rotational symmetry this yields
\begin{equation}
p(\theta) = \frac{1}{2\pi} \mathrm{ce}_0\left(\theta/2,q\right)^2.
\end{equation}
Here $\mathrm{ce}_0$ is the zeroth order Mathieu C function of the first kind~\cite{arfken2005mathematical}, and $q = -1/(2 D_r\gamma)$ is a dimensionless parameter. This distribution is normalized and periodic on $\theta\in[0,2\pi]$, with a peak at $\theta=0$ for $\gamma>0$. We use the standard convention that $\int_0^\pi \mathrm{ce}_0(x,q)^2\mathrm{d}x = \pi$. We will call $p(\theta)= \frac{1}{2\pi} \mathrm{ce}_0\left(\theta/2,q\right)^2$ the Mathieu distribution. This distribution has previously been studied as the solution to some problems in quantum mechanics~\cite{hradil2006minimum,vrehavcek2008experimental}.
The corresponding steering force is given by
\begin{equation}
\mu(\theta) = 4D_r \frac{\partial_\theta\mathrm{ce}_0\left(\theta/2,q\right)}{\mathrm{ce}_0\left(\theta/2,q\right)},
\end{equation}
where $\partial_\theta$ denotes a partial derivative with respect to $\theta$.

\subsection{Pareto Frontier}
For this steering strategy it is possible to evaluate both the average velocity and information rate in closed form using the Mathieu characteristic function $a_0(q)$ and its derivative $a_0'(q)$. To do so, we first note that the Mathieu function $y(x) = \mathrm{ce}_0(x,q)$ satisfies the ODE
\begin{equation}
\underbrace{\left[-\frac{\mathrm{d}^2}{\mathrm{d}x^2} + 2q\cos(2x) \right]}_{=H(q)} y = a_0(q) y,
\end{equation}
where we have defined the operator $H(q)$, which we note is self-adjoint as a result of periodic boundary conditions. Taking a partial derivative with respect to $q$, multiplying both sides by $y$, and integrating over $x$ from $0$ to $\pi$, we obtain
\begin{equation}\label{eq:secondaqeqn}
\int_0^\pi y\frac{\partial H}{\partial q}y\mathrm{d}x + \int_0^\pi yH(q)\frac{\partial y}{\partial q}\mathrm{d}x = a'_0(q)\int_0^\pi y^2\mathrm{d}x + a_0(q)  \int_0^\pi y\frac{\partial y}{\partial q}\mathrm{d}x.
\end{equation}
Using the self-adjoint nature of $H$, we can write 
\begin{subequations}
\begin{align}
\int_0^\pi yH(q)\frac{\partial y}{\partial q}\mathrm{d}x & = \int_0^\pi H(q)y \frac{\partial y}{\partial q}\mathrm{d}x\\
& = a_0(q)\int_0^\pi y \frac{\partial y}{\partial q}\mathrm{d}x.
\end{align}
\end{subequations}
This simplifies Eq.~\eqref{eq:secondaqeqn} to
\begin{equation}
\int_0^\pi y\frac{\partial H}{\partial q}y\mathrm{d}x  = a'_0(q)\int_0^\pi y^2\mathrm{d}x.
\end{equation}
Noting now that $\partial H/\partial q = 2\cos(2x)$, we can simplify and rearrange this equation to yield
\begin{equation}\label{eq:intcosy2result}
\frac{1}{2}a'_0(q) = \frac{\int_0^\pi \cos(2x)y^2\mathrm{d}x}{\int_0^\pi y^2\mathrm{d}x}.
\end{equation}
Since our solution $p(\theta) = y(\theta/2)^2/\pi$, the right hand side of this equation is exactly $\langle \cos\theta\rangle$, so that we have
\begin{equation}
\langle\cos\theta\rangle = \frac{1}{2}a'_0(q).
\end{equation}

We then turn to evaluate the information rate $i$, which we can write in terms of $y(x)$ as
\begin{subequations}
\begin{align}
i/D_r & = \int_0^{2\pi}p(\theta)\left[\partial_\theta\ln p(\theta)\right]^2\mathrm{d}\theta\\
& = \frac{\int_0^\pi (y'(x))^2\mathrm{d}x}{\int_0^\pi y(x)^2\mathrm{d}x}.
\end{align}
\end{subequations}
To evaluate this expression, we return to our original ODE for $y$, multiply both sides by $y$, and integrate from $x=0$ to $\pi$ to obtain
\begin{equation}
\int_0^\pi yy''\mathrm{d}x + a_0(q) \int_0^\pi y^2\mathrm{d}x = 2q\int_0^\pi \cos(2x)y^2\mathrm{d}x.
\end{equation}
Using our result~\eqref{eq:intcosy2result}, we can simplify this to
\begin{equation}
qa_0'(q) - a_0(q) = \frac{\int_0^\pi yy''\mathrm{d}x}{\int_0^\pi y^2\mathrm{d}x}.
\end{equation}
Using integration by parts to show $\int yy'' = -\int y'^2$, we can then relate the right-hand side to the information rate, for which we obtain
\begin{equation}
i/D_r =  a_0(q)-qa_0'(q).
\end{equation}

We thus have a parametric form of the Pareto frontier that can easily be evaluated numerically:
\begin{subequations}
\begin{align}
v/v_0 & = \frac{1}{2} a_0'(q),\\
i/D_r & = a_0(q) - q a_0'(q).
\end{align}
\end{subequations}

\subsection{Approximations to the solution for steering $\mu(\theta)$}\label{sec:vonMises}
As has been noted previously~\cite{hradil2006minimum,vrehavcek2008experimental}, the Mathieu distribution is well-approximated by a von-Mises distribution for both large and small $q$, which correspond to the low and high information limits respectively. The von-Mises distribution takes the form
\begin{equation}
p(\theta) = \frac{1}{2\pi I_0(\kappa)} \exp\left[\kappa\cos(\theta)\right],
\end{equation}
with shape parameter $\kappa$. $I_0(\kappa)$ is the $0$'th order modified Bessel function of the first kind.

As $q\to 0$ (low information) the Mathieu distribution is approximated by a von-Mises distribution with $\kappa = -q$, which leads to a steering strategy
\begin{equation}
\mu(\theta)\approx -\frac{1}{\gamma}\sin\theta.
\end{equation}
With this steering strategy we obtain the scaling
\begin{equation}
v/v_0\approx\sqrt{\frac{i/D_r}{2}}.
\end{equation}
Conversely, as $q\to-\infty$ (high information), the Mathieu distribution is approximated by a von-Mises distribution with $\kappa = \sqrt{-q}$, leading to a steering strategy
\begin{equation}
\mu(\theta)\approx -\sqrt\frac{2D_r}{{\gamma}}\sin\theta,
\end{equation}
and velocity-information scaling
\begin{equation}
v/v_0\approx 1 - \frac{1}{2i/D_r}.
\end{equation}

More generally, we find that the von-Mises ansatz for $p(\theta)$ is a remarkably good approximation to the true optimal solution even at intermediate information rates. This functional form allows us to derive an implicit approximation for the Pareto frontier,
\begin{equation}\label{eq:vonmisesfrontier}
\frac{v}{v_0} = \frac{I_1\left(\frac{i/D_\mathrm{r}}{v/v_0}\right)}{I_0\left(\frac{i/D_\mathrm{r}}{v/v_0}\right)},
\end{equation}
with $I_0$ and $I_1$ the $0$'th and first order modified Bessel function of the first kind

\subsection{Sign-only steering $\mu_{\text{left-right}}(\theta)$}\label{sec:signonly}
For steering, perhaps the simplest solution uses only the sign of $\theta$, adopting a fixed steering rate left or right as required. This could be the case if, for example, the steering force is constrained to only take a single magnitude $|M|$, or if only the sign of $\theta$ can be measured. Using only the sign of $\theta$, the control strategy must take on the following odd functional form:
\begin{equation}
\mu_{\text{left-right}}(\theta)=\begin{cases}
-M, & 0<\theta<\pi\\
M, & \text{else.}
\end{cases}
\end{equation}
We can solve for the resulting steady-state probability distribution, which is
\begin{equation}
p(\theta) = \frac{M\exp\left(\frac{M(\pi-|\theta|)}{D_r+D_c}\right)}{2(D_r+D_c)\left(\exp\left(\frac{M\pi}{D_r+D_c}\right)-1\right)}.
\end{equation}
The velocity and information rate are similarly analytically tractable, and are given by
\begin{subequations}
\begin{align}
v/v_0 & = \frac{M^2\coth\left(\frac{M\pi}{2(D_r+D_c)}\right)}{M^2 + (D_r+D_c)^2},\\
i & = \frac{M^2}{4D_c}.
\end{align}
\end{subequations}
Combining the two to eliminate $M$, we obtain the full pareto frontier
\begin{equation}
v/v_0 = \frac{\coth\left(\frac{\pi}{2}\sqrt{\frac{i}{D_r}}\right)}{1 + \frac{D_r}{i}}.
\end{equation}
For low information, the limiting behavior is
\begin{equation}
v/v_0 \sim\frac{2}{\pi}\sqrt{i/D_r},
\end{equation}
while at high information we have
\begin{equation}
v/v_0 \sim 1 - \frac{D_r}{i}.
\end{equation}
At low information, this has performance only slightly below that
of the full solution (coefficient $2/\pi\approx0.64$ vs. $1/\sqrt{2}\approx0.71$),
while at high information, it needs twice as much information to obtain the same velocity. 

\subsection{Steering without sign $\mu(\left|\theta\right|)$, at low information rate}

We now consider a more constrained case of the continuous steering problem, where the drift $\mu(\theta)$ is required to be an even function of $\theta$. In this case the steady-state distribution no longer takes the Boltzmann form, and there may generally be a net circular flux $J$. The FPE is (setting $D=D_c+D_r$)
\begin{equation}
J = -\mu(\theta)p(\theta) + D p'(\theta).
\end{equation}
Integrating once more reveals that $2\pi J = -\langle\mu\rangle$. Solving for $\mu(\theta)$ as a function of $p(\theta)$, we have
\begin{equation}
\mu(\theta) = -\frac{J}{p(\theta)} + D\frac{p'(\theta)}{p(\theta)}.
\end{equation}
We expand around the low information limit, where $\mu$ and $p(\theta)$ are both constant so that up-gradient velocity is zero. For large $\gamma$ and requiring $\mu(\theta)$ to take the form of a cosine series to enforce evenness, we obtain
\begin{subequations}
\begin{align}
\mu(\theta) & = \pm 2D_r \mp \frac{1}{2\gamma}\cos(\theta).
\end{align}
\end{subequations}
The corresponding velocity and information rate are
\begin{subequations}
\begin{align}
v/v_0& = \frac{1}{16\gamma D_r},\\
i & = \frac{1}{32\gamma^2 D_r},
\end{align}
\end{subequations}
so that the final scaling is 
\begin{equation}
\frac{v}{v_0} = \sqrt{\frac{i}{8D_r}}.
\end{equation}
We can also compute the average velocity perpendicular to the gradient, which we find is equal in magnitude to the up-gradient velocity.

\newpage
\section{Discreteness of the optimal target distribution}

For this section, we consider arbitrary jumping strategies. These are specified by the jump rate $\lambda(\theta)$ and the target jump distributions $q_\theta(\Delta \theta)$. While we are allowing the distributions to vary over $\theta$, we focus on the case of symmetric jump distributions, such that $q_\theta(\Delta \theta) = q_\theta(-\Delta \theta)$. Under this specification, the master equation becomes
\begin{align} \label{eq:master_equation_jumps}
    \frac{dp(\theta)}{dt} = - \lambda(\theta) p(\theta) + \int d\Delta \theta \, \lambda(\theta - \Delta \theta) p(\theta - \Delta \theta) q_{\theta-\Delta \theta}(\Delta \theta) + D_r p''(\theta).
\end{align}
Using the above master equation as a constraint, our optimization problem can be written as the following augmented problem:
\begin{align} \label{eq:master_eq_problem}
    \max_{p, \lambda, (q_\theta)_{\theta}} \int d\theta \, p(\theta) \cos(\theta) - \gamma i \quad \textrm{s.t.} \quad - \lambda(\theta) p(\theta) + \int d\Delta \theta \, \lambda(\theta - \Delta \theta) p(\theta - \Delta \theta) q_{\theta-\Delta \theta}(\Delta \theta) + D_r p''(\theta) = 0 \quad \forall \theta,
\end{align}
where $p$ and $q_\theta$ are subject to normalization constraints, the distributions $q_\theta$ are symmetric, and all choice variables are non-negative.

To invoke some useful results from measure theory, it is convenient to write the target distributions in terms of their corresponding measures. Let $Q_\theta$ be the measure associated to the target distribution at $\theta$, such that
\begin{align}
    dQ_\theta(\Delta \theta) = q_\theta(\Delta \theta) \, d\Delta \theta.
\end{align}
Now, let $\langle\lambda\rangle \coloneqq \int d\theta \, p(\theta) \lambda(\theta)$. Note that $p(\theta) \lambda(\theta)/\langle \lambda \rangle$ is the density of the distribution of locations where jumps are originated. Therefore, the probability of jump sizes averaged over jump sources is captured by the mixture $\bar{Q}$: %
\begin{align}
    \bar{Q} \coloneq \frac{1}{\langle\lambda\rangle} \int d\theta \, p(\theta) \lambda(\theta) Q_\theta.
\end{align}
Let us denote the collection of target measures with ${\bf Q} \coloneqq (Q_\theta)_\theta$. We can write our information rate in terms of $(p, \lambda, {\bf Q})$ as
\begin{align}
    i(p,\lambda, {\bf Q}) = \int d\theta \, p(\theta) \lambda(\theta) \left[ \log\left( \frac{\lambda(\theta)}{\langle\lambda\rangle} \right) + D_{\text{KL}}(Q_\theta \Vert \bar{Q}) \right]
\end{align}
In terms of the target measures, the Lagrangian of the problem \eqref{eq:master_eq_problem} is
\begin{align} \label{eq:augmented_lagrangian}
    {\cal L}_1(p,\lambda,{\bf Q}) =& \int d\theta \, p(\theta) [\cos(\theta)  + D_r \chi''(\theta) + \psi_p] \nonumber \\
    &+ \int d\theta \, p(\theta) \lambda(\theta) \left[ -\gamma \log\left( \frac{\lambda(\theta)}{\langle\lambda\rangle} \right) -\gamma D_{\text{KL}}(Q_\theta \Vert \bar{Q}) + \int dQ_\theta(\Delta \theta) \, \chi(\theta+ \Delta \theta) -\chi(\theta)\right],
\end{align}
where $\psi_p$ is the Lagrange multiplier that enforces the normalization of $p$ and $\chi(\theta)$ enforces the master equation constraint. We now separate the problem of optimization over target measures ${\bf Q}$ and optimization over $(p,\lambda)$.

\subsection{Optimization over target measures}
For fixed $p$ and $\lambda$, we can focus on the elements of the Lagrangian ${\cal L}_1$ that depend on the measures ${\bf Q}$. This inner problem is similar to the one studied in \cite{jung2019discrete} in the context of rational inattention. 

Within ${\cal L}_1$, the only parts that depend on ${\bf Q}$ can be grouped into the following ``inner'' Lagrangian:
\begin{align}
    {\cal L}_2({\bf Q}) \coloneqq \int d\theta \, p(\theta) \lambda(\theta) \left[ \int dQ_\theta(\Delta \theta) \, \chi(\theta+ \Delta \theta) - \gamma D_{\text{KL}}(Q_\theta \Vert \bar{Q}) \right].
\end{align}
Therefore, for a fixed $p$ and $\lambda$, optimizing ${\cal L}_1$ over ${\bf Q}$ is equivalent to solving
\begin{align}
    \max_{{\bf Q}} \, {\cal L}_2({\bf Q}) \quad \text{subject to } Q_\theta \text{ symmetric } \forall \theta.
\end{align}
As we show in Section \ref{subsubsec:augmented_problem}, this problem is equivalent to solving the following augmented problem (for which a symmetric solution always exists):
\begin{align} \label{eq:augmented_inner_problem}
    \max_{{\bf Q}, R} \, {\cal L}_2'({\bf Q}, R),
\end{align}
where now $(Q_\theta)_\theta$ and $R$ are unrestricted probability measures, and we have defined the augmented Lagrangian
\begin{align}
    {\cal L}_2'({\bf Q}, R) \coloneqq \int d\theta \, p(\theta) \lambda(\theta) \left[ \frac{1}{2} \int dQ_\theta(\Delta \theta) \, [\chi(\theta+ \Delta \theta) + \chi(\theta- \Delta \theta)] - \gamma D_{\text{KL}}(Q_\theta \Vert R) \right].
\end{align}
As we will show, at the optimum this problem satisfies $R^* = \bar{Q}^*$, which is necessary for these problems to be equivalent.

Let us first do the optimization in \eqref{eq:augmented_inner_problem} over the measures ${\bf Q}$. Note that this can be done independently for each $\theta$ by solving the problem
\begin{align}
    \max_{Q_\theta} {\cal L}_\theta(Q_\theta), \quad {\cal L}_\theta(Q_\theta) \coloneqq \frac{1}{2} \int dQ_\theta(\Delta \theta) \, [\chi(\theta+ \Delta \theta) + \chi(\theta- \Delta \theta)] - \gamma D_{\text{KL}}(Q_\theta \Vert R).
\end{align}
Using the Donsker-Varadhan variational formula for the KL divergence, this problem can be solved in closed form. The unique maximizer (given $R$) $Q_{\theta,R}^*$ satisfies
\begin{align} \label{eq:Q_optimizer}
    dQ_{\theta,R}^*(\Delta \theta) = \frac{e^{[\chi(\theta + \Delta \theta) + \chi(\theta - \Delta \theta)]/2\gamma}}{Z_R(\theta)} \, dR(\Delta \theta).
\end{align}
Furthermore, the value of our objective function is
\begin{align}
    {\cal L}_\theta(Q_{\theta,R}^*) = \gamma \log\Bigg( \underbrace{ \int dR(\Delta \theta) \, e^{\frac{1}{2\gamma}[\chi(\theta + \Delta \theta) + \chi(\theta - \Delta \theta)]}}_{\eqqcolon Z_R(\theta)} \Bigg).
\end{align}

Using the solution to our optimization of ${\cal L}_\theta$, we can return to the problem \eqref{eq:augmented_inner_problem} and maximize over $R$. Note that
\begin{align}
    {\cal L}_2'({\bf Q}^*_R, R) = \int d\theta \, p(\theta) \lambda(\theta) {\cal L}_\theta(Q^*_{\theta,R}) = \gamma \int d\theta \, p(\theta) \lambda(\theta) \log(Z_R(\theta)).
\end{align}
Therefore, our optimization over $R$ is 
\begin{align} \label{eq:R_problem}
    \max_R \, {\cal L}_2'({\bf Q}_R, R) = \max_R \left\{ \gamma \int d\theta \, p(\theta) \lambda(\theta) \log(Z_R(\theta)) \right\}.
\end{align}
In order to solve problem \eqref{eq:R_problem}, we use the technical result in Section \ref{subsubsec:measure_KKT} on optimization over measures. First, note that the functional ${\cal L}_2'({\bf Q}_R, R)$ is concave, since $R$ enters linearly into the logarithm. We also have the following functional derivative (letting $r(\Delta \theta)$ be the density of $R$):
\begin{align}
    \frac{\delta {\cal L}_2'({\bf Q}_R, R)}{\delta r(\Delta \theta)} = \gamma \int d\theta \, p(\theta)  \lambda(\theta) \frac{e^{[\chi(\theta + \Delta \theta) + \chi(\theta - \Delta \theta)]/2\gamma}}{Z_R(\theta)} \eqqcolon \gamma \langle \lambda \rangle \Psi_R(\Delta \theta).
\end{align}
The function $\Psi_R$ is called the \textit{contact function}, and it determines the support of the optimizer. Using the result in Section \ref{subsubsec:measure_KKT}, we have that there exists a constant $\kappa$ such that, at the optimizer $R^*$,
\begin{align}
    \gamma \langle \lambda \rangle \Psi_{R^*}(\Delta \theta) \le \kappa \quad \forall \Delta \theta, \quad \gamma \langle \lambda \rangle \Psi_{R^*}(\Delta \theta) = \kappa \quad \forall \Delta \theta \in \text{supp}(R^*).
\end{align}
Integrating the second equality against $R^*$ and using Eq. \eqref{eq:Q_optimizer} implies that $\kappa = \gamma \langle \lambda \rangle$. Therefore, our contact conditions are
\begin{align}
    \Psi_{R^*}(\Delta \theta) \le 1 \quad \forall \Delta \theta, \quad \Psi_{R^*}(\Delta \theta) = 1 \quad \forall \Delta \theta \in \text{supp}(R^*).
\end{align}

We can now confirm that this solution satisfies $R^* = \bar{Q}^*$. Using the solution in Eq. \eqref{eq:Q_optimizer}, we see that the density of $\bar{Q}^*$ is
\begin{align} \label{eq:mixture_verification}
    d\bar{Q}^*(\Delta \theta) &= \frac{1}{\langle \lambda \rangle} \int d\theta \, p(\theta) \lambda(\theta) dQ^*_\theta(\Delta \theta) \nonumber \\
    &= \frac{1}{\langle \lambda \rangle} \int d\theta \, p(\theta) \lambda(\theta) \frac{e^{[\chi(\theta + \Delta \theta) + \chi(\theta - \Delta \theta)]/2\gamma}}{Z_{R^*}(\theta)} \, dR^*(\Delta \theta) \nonumber \\
    &= \Psi_{R^*}(\Delta \theta) \, dR^*(\Delta \theta).
\end{align}
Using our contact conditions, we know that $\Psi_{R^*} = 1$ for all $\Delta \theta$ in the support of $R^*$. Thus, we have $\bar{Q}^* = R^*$.

Having characterized the optimal target measures, we can return to the optimization over $p$ and $\lambda$.

\subsection{Optimization over $p$ and $\lambda$}
Now we consider fixed $\chi$ and ${\bf Q}$, and optimize the original Lagrangian \eqref{eq:augmented_lagrangian} over $p$ and $\lambda$. Optimizing over $\lambda$ yields
\begin{align} \label{eq:lambda_foc}
    \frac{\delta {\cal L}_1(p,\lambda,{\bf Q})}{\delta \lambda(\theta)} = p(\theta) \left\{ -\gamma \log\left( \frac{\lambda(\theta)}{\langle\lambda\rangle} \right) -\gamma D_{\text{KL}}(Q_\theta \Vert \bar{Q}) + \int dQ_\theta(\Delta \theta) \, \chi(\theta+ \Delta \theta) -\chi(\theta)\right\} = 0.
\end{align}
Similarly, optimizing over $p$ gives
\begin{align} \label{eq:p_foc}
    \frac{\delta {\cal L}_1(p,\lambda,{\bf Q})}{\delta p(\theta)} =& \cos(\theta) - \gamma \lambda(\theta) \left[\log\left( \frac{\lambda(\theta)}{\langle\lambda\rangle} \right) + D_{\text{KL}}(Q_\theta \Vert \bar{Q}) - 1 \right] + \psi_p + D_r \chi''(\theta) \nonumber \\
    &+ \lambda(\theta) \left[ \int dQ_\theta(\Delta \theta) \, \chi(\theta+ \Delta \theta) -\chi(\theta) \right] = 0.
\end{align}
Combining these two first-order conditions yields the following second-order differential equation for $\chi$:
\begin{align}
    D_r \chi''(\theta) + \psi_p + \cos(\theta) + \gamma \lambda(\theta) = 0.
\end{align}
Integrating over $[-\pi,\pi]$ and using periodicity yields
\begin{align}
    \psi_p = - \gamma \underbrace{\frac{1}{2\pi} \int d\theta \, \lambda(\theta)}_{\eqqcolon \lambda_0}.
\end{align}
Therefore, the differential equation for $\chi$ is fully characterized by $\lambda$:
\begin{align} \label{eq:chi_diff_eq}
    D_r \chi''(\theta) + \cos(\theta) + \gamma [\lambda(\theta) - \lambda_0] = 0.
\end{align}
Additionally, note that for symmetric measures ${\bf Q}$, the first-order condition for $\lambda$ can be written as
\begin{align} \label{eq:lambda_foc2}
    -\gamma \log\left( \frac{\lambda(\theta)}{\langle\lambda\rangle} \right) -\gamma D_{\text{KL}}(Q_\theta \Vert \bar{Q}) + \frac{1}{2}\int dQ_\theta(\Delta \theta) \, [\chi(\theta+ \Delta \theta) + \chi(\theta- \Delta \theta)] -\chi(\theta) = 0
\end{align}
Together, Equations \eqref{eq:chi_diff_eq} and \eqref{eq:lambda_foc2} capture the structure of the optimal $p$ and $\lambda$. We can now put together the ${\bf Q}$ and $(p,\lambda)$ optimizations.

\subsection{Joint optimization} \label{subsec:jump_solution}
Under the optimal measures $Q_{\theta}^*$, following Eq. \eqref{eq:Q_optimizer}, the Donsker-Varadhan result gave us
\begin{align}
    \frac{1}{2} \int dQ_\theta^*(\Delta \theta) \, [\chi(\theta+ \Delta \theta) + \chi(\theta- \Delta \theta)] - \gamma D_{\text{KL}}(Q_\theta^* \Vert \bar{Q}^*)  = \gamma \log\left( Z_{\bar{Q}^*}(\theta) \right).
\end{align}
Comparing with Equation \eqref{eq:lambda_foc2}, we get
\begin{align}
    Z_{\bar{Q}^*}(\theta) = \frac{\lambda(\theta)}{\langle\lambda\rangle} e^{\chi(\theta)/\gamma}.
\end{align}
This implies that our contact function at the optimum is 
\begin{align}
    \Psi_{\bar{Q}^*}(\Delta \theta) = \int d\theta \, p(\theta) e^{[\chi(\theta + \Delta \theta) + \chi(\theta -\Delta \theta) - 2 \chi(\theta)]/2\gamma}.
\end{align}
We are now ready to prove that the optimal target measures are discrete.

\subsection{Discreteness of target support}
\label{sec:discrete-target}

To show discreteness of the distribution of targets, we show that the contact function can only equal 1 at a finite number of points. We do this by showing that it is analytic and non-constant, as has been done to show discreteness of optimal priors \cite{mattingly2018maximizing}. 

Now we evaluate all quantities at the optimum. To simplify notation, we define $\Psi \coloneqq \Psi_{\bar{Q}^*}$. First, note that $\chi$ solves the ODE in equation \eqref{eq:chi_diff_eq}. For solutions with analytic $\lambda$ (which we implicitly constrain our optimization to), this implies that $\chi$ is analytic. Since $e^x$ is an analytic function, it follows that $\Psi$ is analytic in $\Delta \theta$. 

Now, note that
\begin{align}
    \Psi''(0) = \frac{1}{\gamma} \int d\theta \, p(\theta) \chi''(\theta).
\end{align}
Taking the master equation \eqref{eq:master_equation_jumps}, multiplying by $\chi(\theta)$ and integrating yields
\begin{align}
    D_r \int d\theta \, p(\theta) \chi''(\theta) + \int d\theta \, p(\theta) \lambda(\theta) \left[\int dQ_\theta(\Delta \theta) \, \chi(\theta+ \Delta \theta) -\chi(\theta) \right] = 0.
\end{align}
Using the first-order condition for $\lambda$ \eqref{eq:lambda_foc}, we can rewrite the second integral as
\begin{align}
    D_r \int d\theta \, p(\theta) \chi''(\theta) + \gamma \int d\theta \, p(\theta) \lambda(\theta) \left[ \log\left( \frac{\lambda(\theta)}{\langle\lambda\rangle} \right) + D(Q_\theta \Vert \bar{Q}) \right] = 0.
\end{align}
Note that this second term is simply the information rate. Therefore,
\begin{align}
    \Psi''(0) = \frac{1}{\gamma} \int d\theta \, p(\theta) \chi''(\theta) = -\frac{i}{D_r} < 0,
\end{align}
since for finite $\gamma$ the solution achieves a strictly positive information rate. Therefore, $\Psi$ is not constant. Along with analyticity, this means that the contact condition $\Psi(\Delta \theta) = 1$ can only hold at a finite number of points. Thus, the support of the distribution of targets at the optimum must be discrete.

\subsection{Contact function for steering strategies}

As with jumping strategies, the optimal steering problem can also be formulated in terms of an optimization problem over $p$, $\mu$ and $D_c$, where the master equation is enforced through the Lagrange multiplier $\chi(\theta)$. The Lagrangian for the steering problem is
\begin{align}
    {\cal L}_{\text{steering}}(p, \mu, D_c) = \int d\theta \, p(\theta) \left[ \cos(\theta) - \frac{\gamma}{4D_c} [\mu(\theta) - \langle \mu \rangle]^2 + \psi_p \right] + \int d\theta \, \chi(\theta) \left[ (D_r + D_c)p''(\theta) - \frac{d}{d\theta}(\mu(\theta) p(\theta)) \right],
\end{align}
where $\psi_p$ enforces the normalization constraint for $p$. For a fixed $\chi$, the optimization problem can be re-written as
\begin{align}
    \max_{p, \mu, D_c} \int d\theta \, p(\theta) \left\{ \cos(\theta) - \frac{\gamma}{4D_c} [\mu(\theta) - \langle \mu \rangle]^2 + \psi_p + (D_r + D_c) \chi''(\theta) + \mu(\theta) \chi'(\theta) \right\}.
\end{align}
The first-order condition for $\mu$ is
\begin{align}
    \frac{\delta {\cal L}_{\text{steering}}(p,\mu,D_c)}{\delta \mu(\theta)} = p(\theta) \chi'(\theta) - \frac{\gamma}{2D_c} p(\theta) [\mu(\theta) - \langle \mu \rangle] = 0.
\end{align}
Here we can invoke the solution to the problem from Section \ref{subsec:steering_solution}. We have $D_c^*=D_r$, $\langle \mu \rangle = 0$ and $\mu(\theta) = 2D_r \partial_\theta \log(p(\theta))$. Using this in our differential equation for $\chi$ yields
\begin{align}
    \chi'(\theta) = \gamma \partial_\theta \log(p(\theta)) \quad \implies \quad \frac{\chi(\theta)}{\gamma} = \log(p(\theta)) + c_0,
\end{align}
for some constant $c_0$.

Since the contact function for jump strategies in Section \ref{subsec:jump_solution} only depends on the Lagrange multiplier $\chi$, we can evaluate it for the steering case. For jumping strategies with non-symmetric targets $Q_\theta$, the contact function would be
\begin{align}
    \Psi(\Delta \theta) = \int d\theta \, p(\theta) e^{[\chi(\theta + \Delta \theta) - \chi(\theta)]/\gamma}.
\end{align}

With our solution for $\chi$ in the steering case, we can evaluate the contact function to be
\begin{align}
    \Psi(\Delta \theta) = \int d\theta \, p(\theta + \Delta \theta) = 1.
\end{align}

\subsection{Equivalence of the augmented optimization problem} \label{subsubsec:augmented_problem}
Let us consider the problem
\begin{align} \label{eq:original_L2}
    \max_{{\bf Q}} \, {\cal L}_2({\bf Q})& \quad \text{subject to } Q_\theta \text{ symmetric } \forall \theta, \nonumber \\
    {\cal L}_2({\bf Q})& \coloneqq \int d\theta \, p(\theta) \lambda(\theta) \left[ \int dQ_\theta(\Delta \theta) \, \chi(\theta+ \Delta \theta) - \gamma D_{\text{KL}}(Q_\theta \Vert \bar{Q}) \right],
\end{align}
Since we are considering symmetric measures, solving the problem above is the same as solving it for the symmetrized objective
\begin{align} \label{eq:rate_distortion}
    {\cal L}_2^s({\bf Q}) = \int d\theta \, p(\theta) \lambda(\theta) \left[ \frac{1}{2} \int dQ_\theta(\Delta \theta) \, [\chi(\theta+ \Delta \theta) + \chi(\theta- \Delta \theta)] - \gamma D_{\text{KL}}(Q_\theta \Vert \bar{Q}) \right].
\end{align}

We will now prove that the optimization problem with the objective function \eqref{eq:rate_distortion} can be solved with symmetric measures, even when the symmetry constraint is lifted. When the symmetry constraint is lifted, the space of admissible measures is expanded, so if the solution of the expanded problem is symmetric, it will also be a solution of the constrained problem. That is, we now consider the problem
\begin{align} \label{eq:unconstrained_problem}
    \max_{{\bf Q}} \, {\cal L}_2^s({\bf Q}).
\end{align}
Consider an arbitrary collection ${\bf Q}$ (not necessarily symmetric). We will prove that symmetrizing it weakly improves the objective function ${\cal L}_2^s$. Define the symmetrized measures
\begin{align}
    Q^s_\theta \coloneqq \frac{1}{2} (Q_\theta + Q_\theta^-),
\end{align}
where $Q_\theta^-$ is the measure obtained under the transformation $\Delta \theta \to - \Delta \theta$. Clearly these measures satisfy
\begin{align}
    \int dQ_\theta(\Delta \theta) \, [\chi(\theta+ \Delta \theta) + \chi(\theta- \Delta \theta)] = \int dQ_\theta^s(\Delta \theta) \, [\chi(\theta+ \Delta \theta) + \chi(\theta- \Delta \theta)].
\end{align}
Therefore, the first part of the objective is unchanged under symmetrization. Additionally, using the convexity of the KL divergence, we have that
\begin{align}
    D_{\text{KL}}(Q_\theta^s \Vert \bar{Q}^s) \le D_{\text{KL}}(Q_\theta \Vert \bar{Q}) \quad \forall \theta.
\end{align}
This means that symmetrizing the measures weakly improves the objective:
\begin{align}
    {\cal L}_2^s({\bf Q}^s) \ge {\cal L}_2^s({\bf Q}) \quad \forall {\bf Q}.
\end{align}
Therefore, if a non-symmetric solution exists, it can always be symmetrized to weakly improve the objective. We can, then, find our optimal measures by solving the unconstrained problem.

To make progress, we again augment the problem by optimizing over an additional probability measure $R$:
\begin{align} \label{eq:augmented_problem}
    \max_{{\bf Q}, R} \, {\cal L}_2'({\bf Q}, R),
\end{align}
where the augmented Lagrangian is
\begin{align}
    {\cal L}_2'({\bf Q}, R) \coloneqq \int d\theta \, p(\theta) \lambda(\theta) \left[ \frac{1}{2} \int dQ_\theta(\Delta \theta) \, [\chi(\theta+ \Delta \theta) + \chi(\theta- \Delta \theta)] - \gamma D_{\text{KL}}(Q_\theta \Vert R) \right].
\end{align}
Note that this augmented Lagrangian satisfies
\begin{align}
    {\cal L}_2'({\bf Q}, \bar{Q}) = {\cal L}_2^s({\bf Q}).
\end{align}
This means that our search space reaches all possible values of the optimization problem \eqref{eq:unconstrained_problem}. Now, suppose a solution to \eqref{eq:augmented_problem} problem exists that satisfies
\begin{align}
    R^* = \frac{1}{\langle \lambda \rangle} \int d\theta \, p(\theta) \lambda(\theta) Q_\theta^* = \bar{Q}^*.
\end{align}
This would imply that ${\bf Q}^*$ also solves \eqref{eq:unconstrained_problem}. This condition is verified in \eqref{eq:mixture_verification}, so our augmented problem \eqref{eq:augmented_problem} also solves our original problem \eqref{eq:original_L2}.

\subsection{Optimization over measures} \label{subsubsec:measure_KKT}
In this section, we prove a technical result about the support of the solution when optimizing over measures.

\textbf{Lemma.---}
    Let ${\cal X}$ be a compact metric space and ${\cal P}({\cal X})$ be the set of probability measures over ${\cal X}$. Suppose the function $F: {\cal P}({\cal X}) \to \mathbb{R}$ is concave and that there exists $f:{\cal P}({\cal X}) \times {\cal X} \to \mathbb{R}$ such that
    \begin{align} \label{eq:gateaux_derivative}
        \left.\frac{d}{dt} F((1-t) \mu + t \nu) \right|_{t = 0^+} = \int_{{\cal X}} f_\mu(x) \, d(\nu - \mu)(x).
    \end{align}
    Additionally, suppose $f$ is continuous on $x$ for all $\mu$. Then for any $\mu^*$ that solves the problem
    \begin{align}
        \max_{\mu \in {\cal P}({\cal X})} F(\mu)
    \end{align}
    there exists a constant $\kappa$ such that
    \begin{align}
        f_{\mu^*}(x) \le \kappa \quad \forall x \in {\cal X}, \quad f_{\mu^*}(x) = \kappa \quad \forall x \in \text{supp}(\mu^*).
    \end{align}

\begin{proof}
    Fix $\nu \in {\cal P}({\cal X})$ and a maximizer $\mu^*$. Define the function $\phi: [0,1] \to \mathbb{R}$ as
    \begin{align}
        \phi(t) \coloneqq F((1-t) \mu^* + t \nu).
    \end{align}
    By concavity of $F$, the function $\phi$ is concave. Additionally, since $\mu^*$ is a maximizer of $F$, we have that $\phi$ must attain its maximum at $t=0$. For a concave function that attains its maximum at its left endpoint, we must have $\phi'(0^+) \le 0$. By Eq. \eqref{eq:gateaux_derivative}, this implies that
    \begin{align} \label{eq:integral_inequality}
        \int_{{\cal X}} f_{\mu^*}(x) \, d\nu(x) \le \int_{{\cal X}} f_{\mu^*}(x) \, d\mu^*(x) \eqqcolon \kappa.
    \end{align}
    Note that this bound holds for arbitrary measures. In particular, taking the Dirac measure at $x_0$,
    \begin{align}
        \int_{{\cal X}} f_{\mu^*}(x) \, d\delta_{x_0}(x) = f_{\mu^*}(x_0) \le \kappa.
    \end{align}
    Additionally, note that the maximum over $\nu$ of $\int_{{\cal X}} f_{\mu^*}(x) \, d\nu(x)$ is attained when the measure $\nu$ concentrates its mass on the maxima of $f_{\mu^*}$. Since the inequality in Eq. \eqref{eq:integral_inequality} holds for all $\nu$, it must be the case that $\mu^*$ attains this maximum, so it must also concentrate its mass on the maxima of $f_{\mu^*}$. This implies that
    \begin{align}
        \kappa = \int_{{\cal X}} f_{\mu^*}(x) \, d\mu^*(x) = \max_{x \in {\cal X}} f_{\mu^*}(x).
    \end{align}
    Therefore, we must have $f_{\mu^*}(x) = \kappa$ $\mu^*$-almost everywhere. Since $f_{\mu^*}$ is continuous, this implies the stronger result that $f_{\mu^*}(x) = \kappa$ for all $x \in \text{supp}(\mu^*)$.
\end{proof}

The result above generalizes the usual KKT conditions for constrained optimization to the case of measures. Here, the inequality $f_{\mu^*}(x) \le \kappa$ ensures the non-negativity constraint of the measure. For points on the support, the constraint is slack, and the constraint binds for all points where the inequality is strict. 

\newpage

\section{Navigation in three or more dimensions}
In the main text we briefly discussed how the navigation problem and resulting optimal strategies change in three dimensions. Here we provide the details.

\begin{figure}
\centering \includegraphics[width=0.7\textwidth]{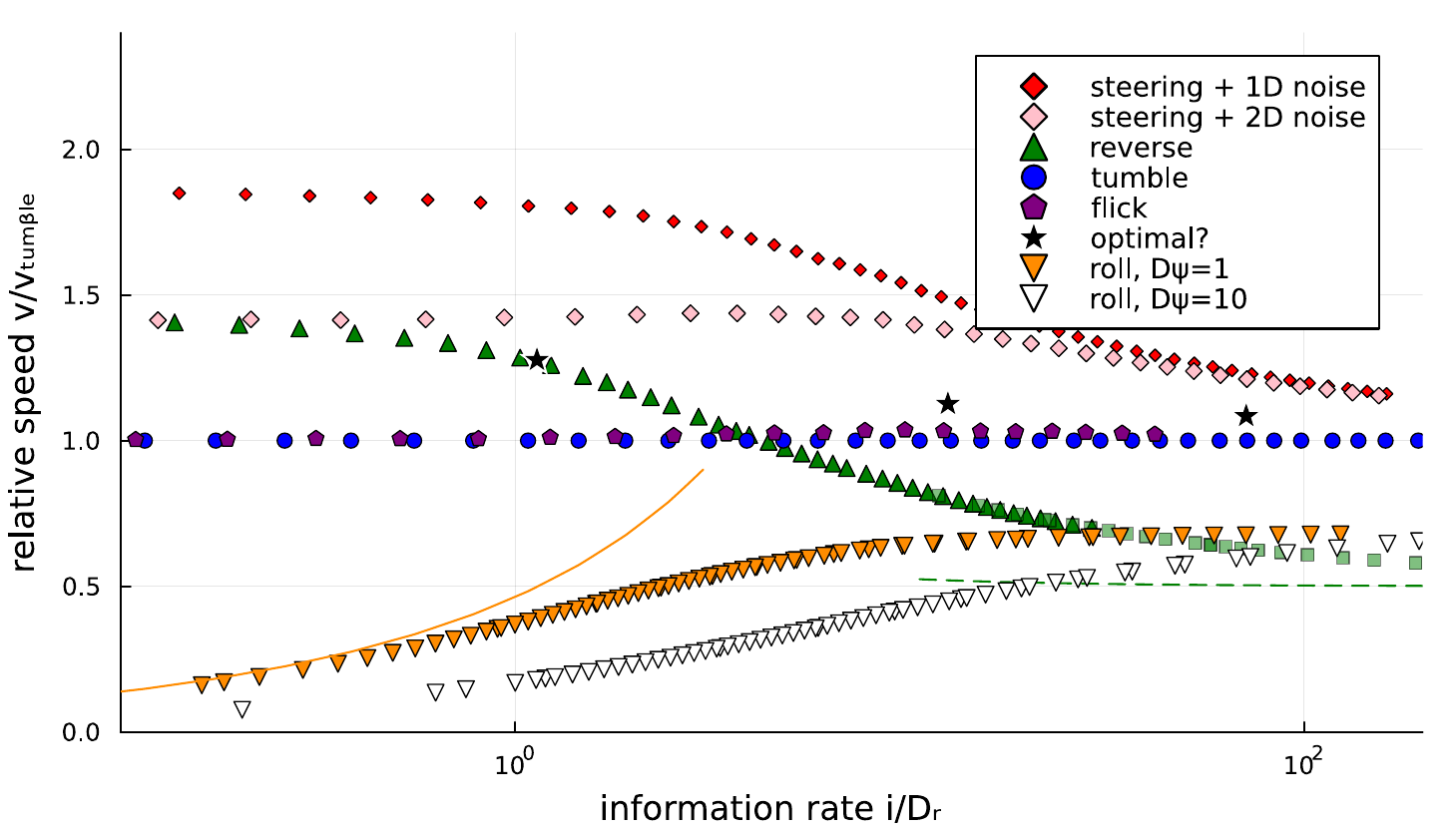}

\caption{Performance of strategies for three-dimensional navigation, relative
to tumbling. Red and pink diamonds show continuous steering with directional information for both ``1D noise'' (red) and ``2D noise'' (pink). Green triangles, blue circles, and purple pentagons respectively show the reverse, tumble, and flick strategies, while black stars indicate the optimal discrete strategy with only scalar information. Inverted triangles show scalar steering with roll diffusion, with $D_\psi=D_r$ (orange) and $D_\psi=10D_r$ (white). Green dashed line shows the asymptotic scaling of the tumble strategy in the high information limit, while the orange solid line shows the conjectured scaling of the unsigned?? steering strategy in the low information limit.}
\label{fig:3D-supp-all}
\end{figure}

\begin{figure}
\centering \includegraphics[width=0.6\textwidth]{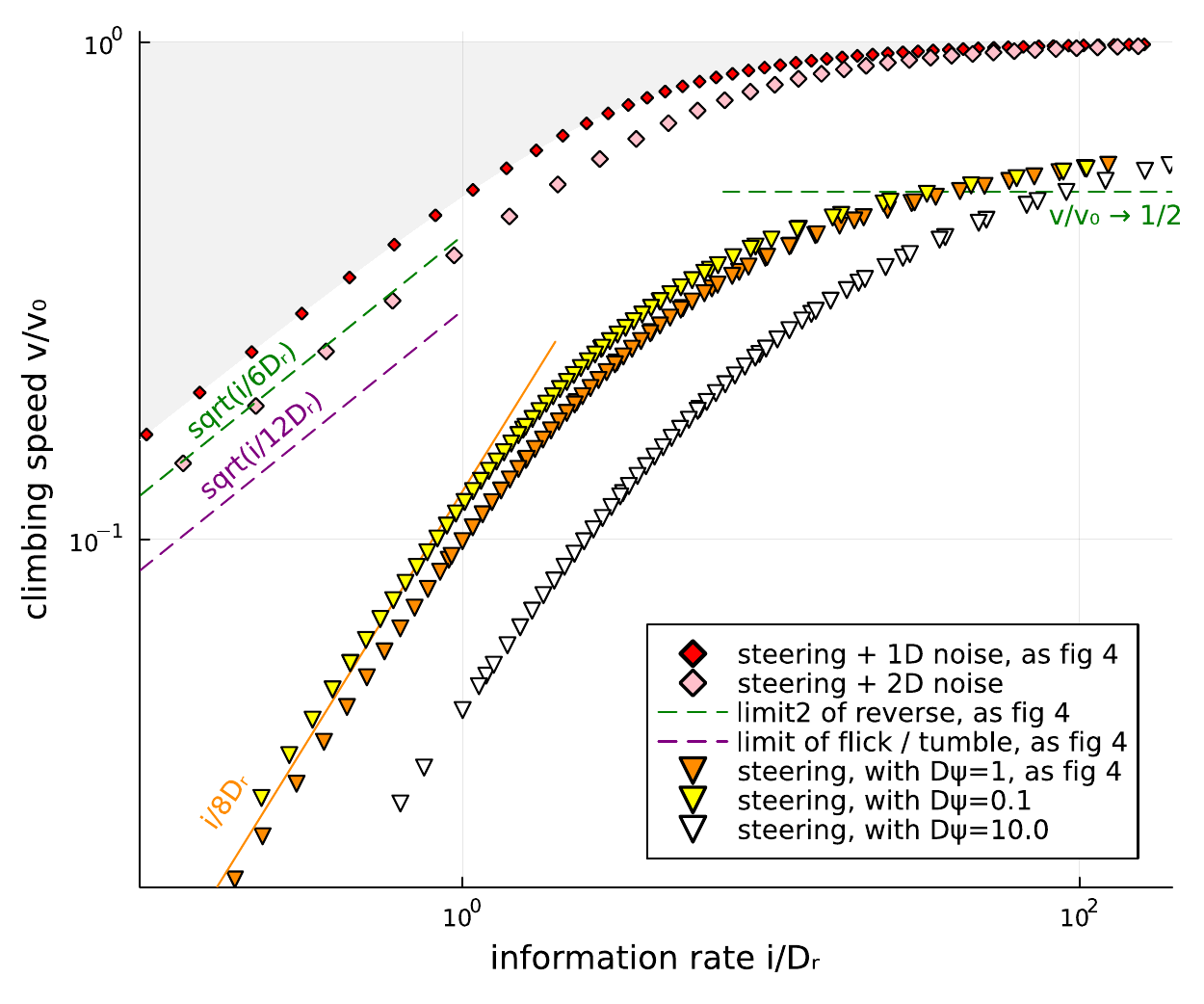}

\caption{More results for 3D steering strategies, with velocity normalized by the maximum swimming speed. Red and pink diamonds show continuous steering with directional information for both ``1D noise'' (red) and ``2D noise'' (pink). Inverted triangles show scalar steering with roll diffusion, with $D_\psi=D_r$ (orange), $D_\psi=0.1D_r$ (yellow), and $D_\psi=10D_r$ (white). Green and purple dashed lines show the low- and high-information scaling for tumble and flick respectively, while the orange solid line shows the conjectured low-information scaling for scalar steering.}
\label{fig:3D-supp-steering}
\end{figure}

\subsection{Simple strategies in $d=3$}
Allowing steering and tumbles, the Fokker-Planck equation on the sphere of possible headings reads 
\begin{equation}
\frac{dp}{dt}=-\vec{\nabla}\cdot(\vec{\mu}p)-\lambda p+\left\langle \lambda\right\rangle +D\nabla^{2}p.
\end{equation}
Assuming radial symmetry $p(\theta,\phi)=p(\theta)$, and allowing complete tumbles, where the new heading is uniform on the sphere, we have:
\begin{align}
\frac{dp(\theta)}{dt} = \underbrace{\int d\theta'\sin\theta'p(\theta')\lambda(\theta')-\lambda(\theta)p(\theta)}_{\text{tumbles}}
+\underbrace{D_r\frac{1}{\sin\theta}\frac{d}{d\theta}\left[\sin\theta\frac{dp(\theta)}{d\theta}\right] }_{\text{diffusion}}.
\end{align}

The simplest kind of steering is only towards or away from the North pole ($\theta=0$). The appropriate controller noise $D_c$ is one-dimensional diffusion along the line of latitude, $\frac{d}{dt}(\sin\theta\,p(\theta)) = D_c \frac{d^2}{d\theta^2}(\sin\theta\,p(\theta))$. This leads to the following Fokker-Planck equation:
\begin{align}
\frac{dp(\theta)}{dt} =-\frac{1}{\sin\theta}\frac{d}{d\theta}\left[\sin\theta\mu(\theta)p(\theta)\right]
+\frac{1}{\sin\theta}\frac{d}{d\theta}\left[(D_r+D_c)\sin\theta\frac{dp(\theta)}{d\theta} + D_c \cos\theta \,p(\theta)\right].
\label{eq:FP-3D-Michael}
\end{align}
This is described as ``1D noise'' on the figures, and gives the red points in figure \ref{fig:3D-maintext} in the main text.
But another, simpler, choice of controller noise is to add another term exactly like the rotational diffusion term $D_r$:
\begin{align}
\frac{dp(\theta)}{dt} =-\frac{1}{\sin\theta}\frac{d}{d\theta}\left[\sin\theta\mu(\theta)p(\theta)\right]
+\frac{1}{\sin\theta}\frac{d}{d\theta}\left[(D_r+D_c)\sin\theta\frac{dp(\theta)}{d\theta}\right].
\label{eq:FP-3D-Matt}
\end{align}
This choice is described as ``2D noise'' on the figures, and the resulting pink points have slightly lower performance.

We could also allow for both reversing the heading at rate $\zeta(\theta)$ and 90 degree flicks at rate $\kappa(\theta)$.
The sink terms are identical, but the source term involves an integral around a circle perpendicular to the original heading vector. Integrating out the azimuthal angle $\phi$, we obtain:
\begin{align}
\frac{dp(\theta)}{dt} =\text{previous}-\underbrace{\zeta(\theta)p(\theta)+\zeta(\pi - \theta)p(\pi-\theta)}_{\text{reverse with rate }\zeta}
-\underbrace{\kappa(\theta)p(\theta)+\frac{1}{2\pi}\int_{0}^{2\pi}d\psi\:\kappa(\theta'(\psi,\theta))p(\theta'(\psi,\theta))}_{\text{flick with rate }\kappa}
\end{align}
where $\theta'(\psi,\theta)=\arccos(-\cos\psi\sin\theta)$.

Figures~\ref{fig:3D-supp-all} and \ref{fig:3D-supp-steering} shows the performance trade-off curves for several of these strategies, computed from numerical solutions to the Fokker-Planck equation. As with the two-dimensional case we find that directed steering is optimal, although now the choice between \eqref{eq:FP-3D-Michael} and \eqref{eq:FP-3D-Matt} gives two different results.
Among strategies not using the direction, The reverse strategy performs well in the low information limit, scaling as $v/v_0\approx \sqrt{i/6D_r}$. Both tumble and flick perform worse, scaling in the low information limit as $v/v_0\approx \sqrt{i/12D_r}$, but both outperform reverse at high information. The reverse strategy reaches a maximum velocity of $v/v_0=1/2$ at high information, while the flick and tumble strategies reach $v/v_0=1$ in the same limit.

\subsection{Information rate for steering in $d$ dimensions}\label{sec:info-derivation-anydim}
While the generalization of the information rate is straightforward for discrete jumps, the generalization for continuous steering is somewhat more involved. Here we derive the information rate for the full $d$-dimensional version of the navigation problem. With only continuous steering, the dynamics for the heading vector $\bm{n}$ on the $d-1$ dimensional sphere are 
\begin{equation}\label{eq:ndsde}
\dot{\bm{n}} = \bm{\mu}(\bm{n}) + \sqrt{2(D_r+D_c)}\bm{\eta}(t),
\end{equation}
with $|\bm{n}|=1$ enforced for all $t$. Note that here we make the choice of ``2D noise'' as described in the previous section. The steering force $\bm{\mu}$ resides in the $d-1$ dimensional space tangent to the $d-1$ sphere at the point $\bm{n}$, as does the random force $\bm{\eta}(t)$. 

We are interested in computing the information rate
\begin{equation}
i = \lim_{\mathrm{d}t\to0} \frac{I[\bm{n};\bm{\Delta_\mathrm{c}n}]}{\mathrm{d}t},
\end{equation}
where 
\begin{equation}
\bm{\Delta_\mathrm{c}n} \coloneq \bm{\mu}(\bm{n})\mathrm{d}t + \sqrt{2D_c\mathrm{d}t}\,\bm{\eta}(t),
\end{equation}
where $\bm{\eta}$ has mean $0$ and covariance given by the $d-1$ dimensional identity matrix $1_{d-1}$. We then rescale this variable to a zero-mean variant,
\begin{equation}
\bm{z} \coloneq \frac{\bm{\Delta_\mathrm{c}n} - \langle \mu\rangle_{p(\bm{n})}\mathrm{d}t}{\sqrt{2 D_c\mathrm{d}t}},
\end{equation}
whose distribution conditional on $\bm{n}$ satisfies
\begin{equation}
p(\bm{z}|\bm{n}) = \mathcal{N}\left[ \sqrt{\frac{\mathrm{d}t}{2D_c}}\left(\bm{\mu}(\bm{n}) - \langle \mu\rangle_{p(\bm{n})}\right),1_{d-1} \right].
\end{equation}
The marginal distribution $p(\bm{z})$ has zero mean. Note that rescaling and shifting a variable has no effect on mutual information, so we can rewrite the information rate as
\begin{equation}
i = \lim_{\mathrm{d}t\to0} \frac{I[\bm{n};\bm{z}]}{\mathrm{d}t}.
\end{equation}
We then rewrite the mutual information as a KL divergence and expand, obtaining
\begin{subequations}
\begin{align}
I[\bm{n};\bm{z}] & = \left\langle D_\mathrm{KL}[p(\bm{z}|\bm{n})\Vert p(\bm{z})]\right\rangle_{p(\bm{n})}\\
& = \left\langle D_\mathrm{KL}[p(\bm{z}|\bm{n}) \Vert \tilde{p}(\bm{z})]\right\rangle_{{p}(\bm{n})} - D_\mathrm{KL}[p(\bm{z})\Vert\tilde{p}(\bm{z})].
\end{align}
\end{subequations}
Here we have defined
\begin{equation}
\tilde{p}(\bm{z}) \coloneq \mathcal{N}( 0,1_{d-1} ).
\end{equation}
The first term is straightforward to evaluate since it is a KL-divergence between two Gaussians with equal covariances, and yields
\begin{subequations}
\begin{align}
\left\langle D_\mathrm{KL}[p(\bm{z}|\bm{n})\Vert \tilde{p}(\bm{z})]\right\rangle_{{p}(\bm{n})} & = \left\langle\frac{1}{2}|\langle\bm{z}\rangle_{p(\bm{z}|\bm{n})}|^2\right\rangle_{{p}(\bm{n})}\\
& = \left\langle\frac{\mathrm{d}t}{4D_c} | \bm{\mu}(\bm{n}) - \langle\bm{\mu}\rangle_{p(\bm{n})}|^2\right\rangle_{{p}(\bm{n})}\\
& = \frac{\mathrm{d}t}{4D_c}\mathrm{Tr}\left[\mathrm{Cov}(\bm{\mu})\right],
\end{align}
\end{subequations}
where the covariance is in the $d-1$ dimensional tangent space.
It then remains to evaluate the second term. Here we define $\bm{m}(\bm{n}) = \langle \bm{z} \rangle_{p(\bm{z}|\bm{n})}$, and note that it is proportional to $\sqrt{\mathrm{d}t}$ and thus small, then note that we can write the marginal distribution as
\begin{subequations}
\begin{align}
p(z) & = \left\langle \tilde{p}\left(\bm{z} - \bm{m}(\bm{n})\right)\right\rangle_{p(\bm{n})}\\
& = \left\langle \tilde{p}\left(\bm{z}\right) \left[ 1 + \bm{z}\cdot \bm{m}(\bm{n}) + \frac{1}{2}\left( [\bm{z}\cdot\bm{m}(\bm{n})]^2 - |\bm{m}(\bm{n})|^2\right) + \mathcal{O}(\bm{m}^3)\right]\right\rangle_{p(\bm{n})}\\
& = \tilde{p}(\bm{z}) \left[1 + 0 + \frac{1}{2}\left(\bm{z}^\top \Sigma_{\bm{m}}\bm{z} - \mathrm{Tr}\Sigma_{\bm{m}}\right) + \mathcal{O}(\mathrm{d}t^{3/2})\right].
\end{align}
\end{subequations}
Here we expanded around $\bm{m}=0$, and kept terms up to first order in $\mathrm{d}t$. We can then evaluate the 2nd KL-divergence term by expanding the integrand around $\mathrm{d}t=0$, which yields
\begin{align}
D_\mathrm{KL}[p(\bm{z})\Vert \tilde{p}(\bm{z})] &  = 0 + \frac{1}{2}\left[\underbrace{\left\langle\bm{z}^\top \Sigma_{\bm{m}}\bm{z}\right\rangle_{\tilde{p}(\bm{z})}}_{\mathrm{Tr}\Sigma_{\bm{m}}} - \mathrm{Tr}\Sigma_{\bm{m}}  \right] + \mathcal{O}(\mathrm{d}t^{3/2})  = 0 + \mathcal{O}(\mathrm{d}t^{3/2}).
\end{align}
Thus after dividing by $\mathrm{d}t$, this contribution to the mutual information is negligible in the small-$\mathrm{d}t$ limit. This concludes our proof, and we have
\begin{equation}
i = \frac{1}{4D_c}\mathrm{Tr}\left[\mathrm{Cov}(\bm{\mu})\right].
\end{equation}

\subsection{Continuous Steering in $d$ Dimensions}
For continuous steering, we analytically obtain a near-optimal strategy in $d$-dimensions with ``2D noise''. When signed information is available, we expect an equilibrium stationary distribution (from the dynamics given by Eq.~\eqref{eq:ndsde}) such that
\begin{equation}
\bm{\mu}(\bm{n}) = (D_r+D_c)\nabla\ln p(\bm{n}).
\end{equation}

Analogous to the von-Mises ansatz in two dimensions, we consider a von-Mises-fisher ansatz here,
\begin{equation}
p(\bm{n}) = \frac{\kappa^{d/2-1}}{(2\pi)^{d/2}1_{d/2-1}(\kappa)} \exp\left(\kappa \bm{r}^\top\bm{n}\right),
\end{equation}
where $\kappa$ is a shape parameter and $\bm{r}$ is the target heading.

The generalization for the velocity is $v/v_0 = \langle \bm{n}\cdot\bm{r}\rangle$, which we can evaluate as
\begin{equation}
v/v_0 = \frac{1_{d/2}(\kappa)}{1_{d/2-1}(\kappa)},
\end{equation}
with $I_m(x)$ denoting the $m$-th order modified Bessel function of the first kind. We can similarly evaluate the information rate, for which we obtain
\begin{equation}
i = \frac{1}{4D_c}\mathrm{Tr}\left[\mathrm{Cov}(\bm{\mu})\right],
\end{equation}
which can be evaluated as
\begin{subequations}
\begin{align}
i/D_r & = (d-1)\kappa\frac{1_{d/2}(\kappa)}{1_{d/2-1}(\kappa)}.
\end{align}
\end{subequations}
Here we have taken $D_c=D_r$, which is optimal by the same arguments we used in two dimensions.

Combining these two results and eliminating $\kappa$, we obtain an analytic performance trade-off curve in arbitrary dimensions, which will be a lower bound on the optimal Pareto frontier:
\begin{equation}
v/v_0 = \frac{1_{d/2}\left(\frac{i/D_r}{(d-1)v/v_0}\right)}{1_{d/2-1}\left(\frac{i/D_r}{(d-1)v/v_0}\right)}.
\end{equation}
Expanding in the low and high information limits this gives
\begin{equation}
v/v_0 \approx\begin{cases} \sqrt{\frac{i/D_r}{d(d-1)}} & i/D_r\ll 1,\\
1 - \frac{(d-1)^2}{2i/D_r} & i/D_r\gg1.\end{cases}
\end{equation}

\subsection{Undirected Steering in 3 Dimensions}
\label{sec:roll-steer-3D}

We then turn to the question of how the even steering solutions in 2 dimensions should be generalized to 3 dimensions. Consider the degree of deviation of the current heading $\bm{n}$ from the target heading $\bm{r}$, which can be quantified by the angle $\theta$ between them such that $\bm{n}\cdot\bm{r}=\cos\theta$. To parameterize $\bm{n}$ we require two angles, $\theta$ (the angle between the heading and the target), and $\phi$ the perpendicular angle with respect to which we expect the problem to be symmetric. Now suppose that the strategy $\bm{\mu}$ can only depend on the value of $\bm{n}\cdot\bm{r}$, or equivalently on $\cos\theta$. What is the optimal strategy under this constraint?

In two dimensions the agent was pinned to a plane, and could thus choose to steer left or right even without knowledge of which direction pointed towards the target direction. In three dimensions, however, this no longer makes sense, since the agent can roll freely around the heading vector $\bm{n}$. We thus treat the steering capabilities of the agent as applying a steering force in a vector direction $\bm{f}$ which is perpendicular to $\bm{n}$, such that $\bm{\mu}=\bm{f} \,g(\theta)$ for some scalar function $g(\theta)$ which we require to be even.

\begin{figure}
\centering \includegraphics[width=0.4\textwidth]{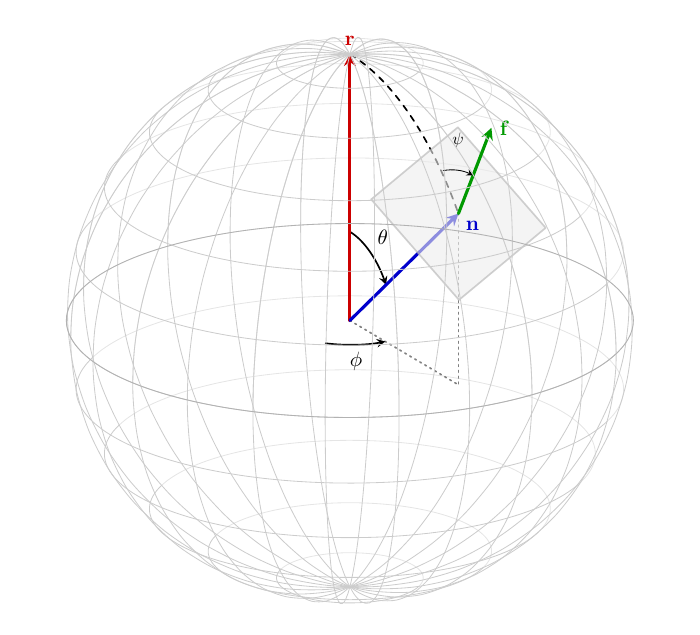}
\caption{Schematic to illustrate the vectors and angles involved in 3D unsigned steering.}
\label{fig:3d_diagram}
\end{figure}

To make progress, we define an additional ``facing'' vector $\bm{f}$ which is perpendicular to $\bm{n}$, and is the direction in which the agent can steer. We will quantify $\bm{f}$ using the angle $\psi$ between $\bm{f}$ and the optimal steering force direction. We would then write the steering force as $\bm{\mu}=\bm{f} \,g(\theta)$. We will describe the dynamics of this steering vector using the angle $\psi$ between $\bm{f}$ and the tangent vector pointing directly towards the north pole. We assume this angle also undergoes rotational diffusion with a diffusion coefficient $D_\psi$. Figure~\ref{fig:3d_diagram} illustrates the geometry of the problem.

Carefully keeping track of the relevant geometric and trigonometric factors, we find that three angles in the problem evolve according to the following stochastic differential equations:
\begin{subequations}
\begin{align}
\dot{\theta} & = -g(\theta)\cos\psi + (D_r+D_c)\cot\theta+ \sqrt{2(D_c+D_r)}\eta_\theta(t),\\
\dot{\phi} & = -g(\theta)\sin\psi/\sin\theta + \frac{1}{\sin\theta}\sqrt{2(D_c+D_r)}\eta_\phi(t),\\
\dot{\psi} & = g(\theta)\sin\psi\cot\theta + \sqrt{2D_\psi}\eta_\psi(t).
\end{align}
\end{subequations}
Here we have applied the controller noise $D_c$ to both the $\theta$ and $\phi$ directions, and treated heading diffusion on the sphere and roll diffusion as independent.

The associated Fokker-Planck equation is
\begin{equation}
\begin{aligned}
\frac{\partial}{\partial t}p(\theta,\phi,\psi) = & -\frac{1}{\sin\theta}\partial_\theta\left[-g(\theta)\cos\psi\sin\theta \, p(\theta,\phi,\psi) \right] -\frac{1}{\sin\theta}\partial_\phi\left[ -g(\theta)\sin\psi \, p(\theta,\phi,\psi) \right] \\
& - \partial_\psi\left[g(\theta) \cot\theta\sin\psi\, p(\theta,\phi,\psi)\right] + D_\psi\partial_\psi^2 p + (D_r+D_c)\left[\frac{1}{\sin\theta}\partial_\theta\left(\sin\theta\,\partial_\theta p\right) + \frac{1}{\sin^2\theta}\partial_\phi^2p \right].
\end{aligned}
\end{equation}

We then seek to optimize the objective
\begin{equation}
\mathcal{L} = \langle\cos\theta\rangle - \gamma  \frac{1}{4D_c}\mathrm{Tr}\left[\mathrm{Cov}(\bm{\mu})\right]
\end{equation}
with respect to $D_c$ and the function $g(x)$.

We will derive the scaling in the low information limit using a series expansion. In the low information limit we have $g_0=0$ and $p_0 = 1/8\pi^2$. We now expand around this limit in a small parameter $\epsilon\propto1/\gamma$, as
\begin{subequations}
\begin{align}
p & = \frac{1}{8\pi^2} + \epsilon p_1 + \mathcal{O}(\epsilon^2),\\
g & = 0 + \epsilon g_1(\theta) + \mathcal{O}(\epsilon^2).
\end{align}
\end{subequations}
We require that $g_1$ be an even function of $\theta$.

We will also assume $p$ is independent of $\phi$ so that the FPE can be simplified to
\begin{equation}
\begin{aligned}
\frac{\partial}{\partial t}p(\theta,\psi) = & -\frac{1}{\sin\theta}\partial_\theta\left[-g(\theta)\cos\psi\sin\theta \, p(\theta,\psi) \right] - \partial_\psi\left[g(\theta) \cot\theta\sin\psi\, p(\theta,\psi)\right] + D_\psi\partial_\psi^2 p + (D_r+D_c)\frac{1}{\sin\theta}\partial_\theta\left(\sin\theta\,\partial_\theta p\right) .
\end{aligned}
\end{equation}

At first order, we will have
\begin{equation}
\begin{aligned}
0 = & D_\psi\partial_\psi^2 p_1 + (D_r+D_c)\frac{1}{\sin\theta}\partial_\theta\left(\sin\theta\,\partial_\theta p_1\right)
  -\frac{1}{\sin\theta}\partial_\theta\left[-g_1\cos\psi\sin\theta \, p_0 \right] - \partial_\psi\left[ g_1 \cot\theta\sin\psi\, p_0\right].\\
= & D_\psi\partial_\psi^2 p_1 + (D_r+D_c)\left[\partial_\theta^2p_1 + \cot\theta\,\partial_\theta p_1\right]
 + \cos\psi\,p_0\,\partial_\theta g_1.
\end{aligned}
\end{equation}
This functional form strongly suggests the ansatz $p_1(\theta,\psi) = T(\theta)\cos(\psi)$. Any higher harmonics in $\psi$ could exist transiently, but should decay to zero at steady state due to the $\cos\psi$ forcing. This ansatz allows us to write 
\begin{equation}
0 = -D_\psi\cos\psi T(\theta) + \cos\psi (D_r+D_c)\left[T''(\theta) + \cot\theta\,T'(\theta)\right] + \frac{1}{8\pi^2}\cos\psi g_1'(\theta).
\end{equation}
Dividing through by $\cos\psi$ then gives an ODE for $T(\theta)$,
\begin{equation}
0 = -D_\psi T(\theta) +(D_r+D_c)\left[T''(\theta) + \cot\theta\,T'(\theta)\right] + \frac{1}{8\pi^2}g_1'(\theta).
\end{equation}

We can then proceed to evaluate the velocity and information rate. We begin with the information rate, which is
\begin{equation}
\begin{aligned}
i & = \frac{1}{4D_c} \mathrm{Tr}\left[\mathrm{Cov}(\bm{\mu})\right]\\
& = \frac{1}{4D_c} \langle g^2\rangle_{p(\theta,\phi,\psi)}\\
& = \frac{1}{4D_c} \int\int\int \epsilon^2 g_1(\theta)^2 p_0(\theta,\phi,\psi)\sin\theta\mathrm{d}\theta\mathrm{d}\phi\mathrm{d}\psi + \mathcal{O}(\epsilon^3)\\
& = \frac{\epsilon^2}{8D_c}\int g_1(\theta)^2 \sin\theta \mathrm{d}\theta + \mathcal{O}(\epsilon^3).
\end{aligned}
\end{equation}
Since $\sin\theta>0$ on $(0,\pi)$, so long as $g_1$ is not identically zero for all $\theta$ the information rate will be second order in $\epsilon$.
We similarly evaluate the velocity as
\begin{equation}
\begin{aligned}
v/v_0 & = \langle \cos\theta\rangle\\
& = \epsilon\int\int\int\cos\theta p_1(\theta,\phi,\psi)\sin\theta\mathrm{d}\theta\mathrm{d}\phi\mathrm{d}\psi + \mathcal{O}(\epsilon^2)\\
& = 2\pi\epsilon \left(\int \cos\psi \mathrm{d}\psi\right)\left(\int T(\theta)\cos\theta\sin\theta \mathrm{d}\theta\right) + \mathcal{O}(\epsilon^2)\\
& = 0 + \mathcal{O}(\epsilon^2).
\end{aligned}
\end{equation}
Unlike in two dimensions, the up-gradient velocity is exactly zero to first order in $\epsilon$, while the information rate is non-zero at second order in $\epsilon$. Thus $v\sim\sqrt{i}$ scaling is impossible, and the low information velocity-information scaling can be at best $v\sim i$.

\end{document}